\documentclass[pre,showpacs,showkeys,floatfix, notitlepage,nofootinbib]{revtex4-1}

\usepackage{amsmath}
\usepackage{amsfonts}
\usepackage{graphicx}
\usepackage{bm}
\usepackage{amssymb, amsthm}
\usepackage{mathrsfs}
\usepackage{pstricks,pst-node}
\usepackage{amsxtra}
\usepackage{bbm}
\usepackage{paralist}
\usepackage{color}
\usepackage{soul}
\usepackage{tabularx} 
\usepackage{arydshln} 

\DeclareMathAlphabet{\mathbfi}{OT1}{cmr}{bx}{it}
\DeclareMathAlphabet{\mathpzc}{OT1}{pzc}{m}{it}

\newcommand{\eqa}{\begin{eqnarray}}
\newcommand{\eeqa}{\end{eqnarray}}
\newcommand{\beq}{\begin{equation}}
\newcommand{\eeq}{\end{equation}}

\newtheorem{theorem}{Theorem}

\newtheorem{proposition}[theorem]{Proposition}

\newcommand{\benumerate}{\begin{enumerate}}
\newcommand{\eenumerate}{\end{enumerate}}

\newcommand{\beit}{\begin{itemize}}

\newcommand{\enit}{\end{itemize}}

\usepackage{color}

\usepackage{soul}

\begin{document}

\title{  On solutions to a novel non-evolutionary integrable 1+1 PDE} 

\author{
Francesco Giglio$^{\; 1)}$\footnote{{\tt email: francesco.giglio@glasgow.ac.uk}}, 
Giulio Landolfi$^{\; 2)}$\footnote{{\tt email: giulio.landolfi@le.infn.it, giulio.landolfi@unisalento.it}}, 
Luigi Martina$^{\; 2)}$\footnote{{\tt email: luigi.martina@le.infn.it, luigi.martina@unisalento.it}}
}

\affiliation{$^{1)}$ School of Mathematics and Statistics, University of Glasgow, Glasgow, UK }
\affiliation{$^{2)}$ Dipartimento di Matematica e Fisica ``Ennio De Giorgi" Universit\a`a del Salento \\
and I.N.F.N. Sezione di Lecce, via Arnesano I-73100 Lecce, Italy}

\date{\today}

\begin{abstract}
 We investigate real solutions of a C-integrable non-evolutionary  partial differential equation  in the form of a  scalar conservation law where 
 the flux density depends both on the density and  on its  first derivatives  with respect to the local variables.
 By performing a similarity reduction dictated by one of its local symmetry generators, 
a nonlinear ordinary differential equation arises that is connected to the  Painlev\'e III equation. 
 Exact solutions are secured and described provided a constraint holds among the coefficients of the original equation.
In the most general case, we pinpoint the generation of additional singularities by numerical integration.  
Then, we discuss the evolution of given initial profiles. Finally, we mention aspects concerning rational solutions with a finite number of poles.
\end{abstract}

 
\maketitle

 \section{Introduction}

We address the study of the nonlinear partial differential equation  (PDE)
 \begin{equation}
\label{veq}
\partial_t v + \partial_{x} \left\{ \frac{1}{c_{1} v + c_{3} \sigma} \left[c_{2} v^{2} + c_{4} \sigma^{2} + \sigma \eta \left(c_{1} \partial_{t} v + c_{2} \partial_{x} v\right) \right] \right\} = 0
\end{equation}
where $v=v(x,t)$ is a real function of the two real independent variables $x$ and $t$, and $\eta$, $\sigma$ and the $c_j$'s are real constants. The equation has been introduced recently in \cite{giglio} in connection with the problem of describing fluid systems of volume $v$ whose behavior at pressure $P=x/t$ and temperature $T=1/t$ deviates from the 
van der Waals equations of state, the parameter  $\eta$ being there a small positive real parameter  identifying the inverse of the number of fluid molecules.
  Equation \eqref{veq} extends case studies that are very familiar to researchers in applied mathematics and physics, above all the Bateman-Burgers equation $ \partial_t u+u  \partial_x u=\eta \partial_{xx}u$ \, \cite{bateman,burgers1, burgers2} whose structure is recovered when $c_1=0$. 
In fact, Equation \eqref{veq}  collocates itself naturally  within equations that generalise Kolmogorov-Petrovsky-Piskounov reactive-diffusive \cite{kolmogorov}
 or nonlinear Fokker-Planck models \cite{fokker}, viz. $ \partial_t v+D \partial_{xx}v=f(v)$ and $\partial_t v+D \partial_{xx}v=\partial_x (F v)$, being $D$ the diffusion coefficient. 
 It is known that these equations find several applications in a number of fields, ranging from condensed matter and liquid crystals physics to population dynamics and social sciences (see e.g.   \cite{frank, murray, furioli}), and incorporate  various phenomena, such as phase transitions, travelling-wave solutions, shock waves,  anomalous diffusion, and so forth. 

Owing to the rather general conditions behind its introduction,  one can expect that even the PDE  \eqref{veq} may be relevant, or at least of use in the complex systems domain in a wider perspective,  not limited to the early motivation in \cite{giglio}.  Indeed,  PDE  \eqref{veq} originates  merely from two plain requirements:  

\begin{enumerate}[i)]
\item the insertion within a scalar conservation law
 \beq
\partial_t v = \partial_x \varepsilon  \,\, 
\label{PDE conservation law}
\eeq
of  a simple nonlocal expansion of the flux density
\begin{equation}
\varepsilon =  \varepsilon_{0}(v) + \eta \,\varepsilon_{1}(v) \,\partial_x v + \eta \,\varepsilon_{2}(v) \, \partial_t v + O(\eta^{2}) + g(t) \,\, , 
\label{internal energy expansion}
\end{equation}
where $\varepsilon$ is assumed be uniquely determined by a function $v=v(x,t)$, depending in its turn  on two local variables $x$ and $t$, 
and  $\eta$ is a small expansion parameter; 
\item the successive singling out of those dynamics that are compatible with the linearisability requirement of the  PDE resulting from \eqref{PDE conservation law}-\eqref{internal energy expansion}, via the Cole-Hopf transform 
\beq
 v(x,t)=\eta \,\sigma \, \partial_x \ln\varphi(x,t)\,\, , 
 \label{Cole Hopf v}
 \eeq
into a linear PDE for the  potential function $\varphi(x,t)$, being   $\sigma$ a nonvanishing real parameter. 
 \end{enumerate} 
  
Remark that, contrary to the models previously mentioned,  the C-integrable \cite{calogero} equation \eqref{veq} is actually not of evolutionary type,  like $\partial_t v=G(x,t,v,\partial_x v,\partial_{xx} v,...)$ for which many general  results are available in the literature and  can be exploited \cite{vinogradov, olver, svinolupov, sokolov, abramenko,lagno}.   For instance, it is known that a PDE of the form $\partial_t v=\partial_{xx}v+f(v,\partial_x v)$  can be mapped into the  {\em heat} equation or the Bateman-Burgers equations  (see, e.g., Proposition 4.3 in \cite{vinogradov}, pp. 170). But from results in  \cite{giglio jpa} regarding the symmetry properties one  realises that no local transformations can map   \eqref{veq} into these 
 equations. Notwithstanding, the model can be linearised through the Cole-Hopf transformation  
\eqref{Cole Hopf v} with real-valued parameter  $\sigma$ and potential $\varphi$, whose actual value and meaning may be suggested from the specific problem\footnote{In the real fluid case dealt with in \cite{giglio}, for instance,  $\sigma$ turns out to be connected to the universal gas constant 
(or equivalently to the Boltzmann thermodynamic constant
\cite{giglio jpa}) while $\varphi$ plays the role of the statistical partition function. 	}.
As a result of requirements i) and ii), function $\varphi$ fulfils a   linear partial differential equation that, interestingly, turns out to be an "interpolation" of the \emph{heat} and \emph{Klein-Gordon} type equations\footnote{The connections between statistical field models and the \emph{heat} or Klein-Gordon  equations for the underlying partition function  have been pointed out in literature for archetypical complex systems such as the Curie-Weiss model for spins, the van der Waals model for real gases and generalisations, and the Maier-Saupe model and biaxial generalisations for nematic liquid crystals \cite{zagrebnov, choquard, barra, barra 2, brankov,dematteis, dematteis2}. 
The  application of a Cole-Hopf type transform to the 1+1 \emph{heat} and \emph{Klein-Gordon} equations leads to nonlinear equations in the Burgers hierarchy, that is  viscous scalar conservation laws possessing constant viscous central invariant \cite{arsie}.}:
 \begin{equation}
\label{phieq}
\underbrace{ \eta\,c_{3} \,\partial_t \varphi +\eta^{2} \,c_{2} \,\partial_{xx}\varphi }_{\text{heat-type}}\,\, +\,\,
\underbrace{\eta^{2}\, c_{1} \,\partial_{xt}\varphi  + c_{4} \varphi }_{\text{KG-type}}=0 \,\, . 
\end{equation}

Similarly to the Bateman-Burgers case and other C-integrable PDEs, even though there is the possibility to tackle the problem of analysing its solutions by taking great advantage from its linearisability and from a successively nicely posed initial value problem, there are reasons to not overlook the direct study of Equation \eqref{veq}. For instance, the dynamical  emergence of some peculiar features, such as singularities or shock-type dynamics, may be more perspicuous. Further,  the way one can benefit from known results to generate new ones by resorting to symmetry group techniques is affected. The Cole-Hopf transformation is indeed non-local and the spectrum of local symmetry generators of \eqref{veq} and those of its $\varphi$-potential formulation differ,  and accordingly the explicit form of the associated invariants. 
Moreover, similarly to what it has been argued for the Bateman-Burgers equation, issues can be raised in respect to the problem of giving appropriate boundary conditions, due to the nature of the Cole-Hopf transformation  \cite{kevorkian, maleka}.  Finally, 
once the expansion \eqref{internal energy expansion} is assumed, but the linearisability requirement for 
\eqref{PDE conservation law} is relaxed, models would outcome whose investigation will advantage of what is already apprehended about \eqref{veq}. 

 Because of the aforementioned reasons,  we are going to look for basic distinguishing attributes  exhibited by solution to \eqref{veq},  by paying attention, in particular, to  the presence of singularities.  To do this, we exploit the knowledge of the symmetries identified in \cite{giglio jpa}, so to perform a suitable reduction leading to a nonlinear ordinary differential equation (ODE). Such an ODE describes the solutions on the orbits of a one-dimensional symmetry subgroup, generated by the simultaneous Galilei and scaling transformation for the $x,t,v $ variables, and it turns out to be connected with the Painlev\'e III equation. In the presence of a certain constraint on  the coefficients $c_j$ in \eqref{veq}, solutions can be given in closed form, and their features are discussed. In the most general case,  one can proceed by pursuing specific strategies, such as numerical integration or the carrying  of a Painlev\'e test, to understand relevant characteristics, such as the dynamics of poles.

The outline of the paper is as follows. In Section \ref{section II} we introduce the problem of solving Equation \eqref{veq} by selecting some  natural simplifications, driven by conditions on the parameters. In Section \ref{section similarity reductions} we focus on solutions pertinent to the orbits of the local symmetry generators resulting for the equation from a symmetry group theoretical approach, by first arguing on what the two simplest generators imply and then deriving the nonlinear ODE describing similarity solutions pertinent the remaining generator.
In Section \ref{section Delta 0} we determine the solutions to  the obtained nonlinear ODE, once a given condition is satisfied by coefficients $c_j$, and clarify their features. 
Implications for the corresponding functions $v(x,t)$ solving Equation \eqref{veq} are  in Section  \ref{section Delta 0 v(x,t)}.
The most general case is considered in Section \ref{section Delta gen} where, also bearing in mind findings of Section
\ref{section Delta 0}, we explore the consequences as regards the properties of the solutions to \eqref{veq}, and their pole and stationary points dynamics. In Section \ref{section poles} 
we perform remarks concerning the analysis of pole dynamics implied for rational solutions involving multiple poles. 
In Section \ref{section conclusions} we  summarise the results of our study. Finally,  Appendix \ref{appendice 2} contains proofs for  Propositions  stated  in Section \ref{section Delta 0}. 

\section{Properties of Equation (1) -- Some subclasses of solutions} 
\label{section II}
As we already mentioned, Equation \eqref{veq} represents a generalisation of known diffusive/dissipative models. Several comments can be straightforwardly made in respect to some of its solutions when particular conditions hold on coefficients or when specific dependences on independent variables are  sought. Below, we pay attention to the first prospect, the second one being dealt with in Section \ref{section similarity reductions}.

\subsection{Remarks on subcases implied by the vanishing of coefficients $c_j$}
\label{subsection cj 0}
 
The various real constants present in  \eqref{veq} can assume arbitrary values in general, and depending on the problem to which the model would refer they may acquire particular meaning and major roles under distinct dynamical regimes\footnote{ For example, in the discussion in \cite{giglio}:  i)  setting $\sigma$ to minus the universal gas constant guarantees that at high temperatures $T=t^{-1}$ and pressures $P=x/t$ an ideal gas behavior with a core-volume term can be matched with $\varphi$ being just the associated statistical partition function, and in this regime effects of constants $c_j$ can be neglected; ii) $c_4/c_1=a/\sigma^2$, where $a$ denotes  
the  mean-field parameter $a$ entering the van der Waals equation of state $(v-b)(P+a v^{-2})=RT$ for  real gases 
(see also \cite{barra});  iii)
 sign ansatz on the structural constants $c_j$ are to be  considered to guarantee the reaching of critical regime with a  real fluid behavior and to avoid that the partition function so identified in this description diverges. }.
By acting on the coefficients $c_j$, reductions can be performed on Equation \eqref{veq} that significantly affect the differential problem.
This immediately eventuates in the following examples.
\vskip 0.2cm 
\noindent \textit{Example 1: \underline{ $c_1=0$.} }
 One ends up into a Bateman-Burgers equation structure, which has originally introduced in \cite{bateman,burgers1} and is one among the most investigated canonical integrable nonlinear partial differential equations. With $c_1=0$, Equation \eqref{veq} becomes indeed  the nonhomogeneous nonlinear heat equation 
 \beq
\partial_t v+\frac{2c_2}{\sigma c_3} \, v\, \partial_x v+\eta\frac{c_2}{c_3} \partial_{xx}v=0 \,\, 
\label{IEW burgers}
\eeq
with the thermal diffusion coefficient $-\eta c_2/c_3$. Equation \eqref{IEW burgers} can be converted to a linear diffusion equation by Cole-Hopf transform, as it is widely renowned.
\vskip 0.2cm

\noindent  \textit{Example 2:  \underline{$c_2=c_4=0$.}}
Equation \eqref{veq} becomes
 \begin{equation}
\partial_t v +  \sigma \, \eta \,c_{1} \, \partial_{x} \left( \frac{1}{c_{1} v + c_{3} \sigma}   \partial_{t} v    \right)  = 0\,\,.
\label{veq c2 c4 zero}
\end{equation}
We would like to point out that \eqref{veq c2 c4 zero} can be promptly integrated to give
\beq
   v = \sigma   \eta \, \, \partial_x \ln \left[ \beta (x) + \alpha (t) e^{-\frac{c_3}{c_1\eta} x } \right]   \,\, , 
\label{generalized ideal gas}
\eeq 
being $\alpha(t)$ and $\beta(x)$ arbitrary functions of their variables.  It is thence worth to pay attention on the case $c_2=c_4=0$ because, due to the simultaneous presence of two arbitrary functions, a number of examples could be given  and discussed. We notice, in particular, that simple pole solutions are allowed even on scale other than that induced by the product $\eta \sigma$, as opposite to what results from solutions \eqref{sol stazionarie pm} (even when one among $c_2$ or $c_4$ is zero). Precisely, solutions $v=\eta \sigma \gamma/x$ would follow through the choice $\alpha(t)=0$ and
$\beta(x)=x^\gamma$, for some non-zero constant $\gamma$.\footnote{Among possible noticeable  examples, remaining within the original context in which Equation \eqref{veq} was derived, such choice would enable in principle to account for the ideal gas equation of state. However, keeping the values of $\sigma<0$ and $\eta>0$ identified in \cite{giglio}, the simple pole term $\eta\sigma/x$ in \eqref{sol stazionarie pm} would have instead wrong negative sign and physical scale.} 
 
 \vskip 0.2cm

\noindent  
\textit{Example 3:  \underline{$c_3=c_4=0$.} }
In this case one ends up into a Riccati type equation:
 \begin{equation}
\sigma \, \eta \,c_{1} \, \partial_{\tau} v(\tau,t) =f_1(\tau) \, v(\tau,t) -v(\tau,t)^2  \,\,, \qquad \tau=c_1 x -c_2 t \,\, ,
\label{veq c3 c4 zero}
\end{equation}
where $f_1$ is an arbitrary function  of $\tau$.  Solutions to \eqref{veq c3 c4 zero} can be thus given as 
\beq
v(\tau,t)=c_1 \eta \sigma \partial_\tau \ln \left[ f_2(t) +\int^\tau e^{ \frac{1}{c_1 \eta \sigma} \int^{\tilde{\tau}}  f_1(\chi) d\chi} d\tilde{\tau}
\right]
\eeq
being $f_2(t)$ an arbitrary function of variable $t$.

\subsection{The limit $\eta\to0$ -- weak solutions} 
For the particular choice $c_{1} = 0$, the model equation \eqref{veq} is nothing but the Bateman-Burgers equation, describing the propagation of nonlinear waves in regime of small viscosity. In the inviscid limit $\eta\to 0$ weak solutions
are thence expected indicating a non trivial complex behavior for the associated physical system, such as a phase transition \cite{moro, choquard, barra 2, brankov}.
 A similar picture is expected to hold as well when $c_1\neq0$. Indeed, if $\eta$ is vanishing the implicit solution form 
 \beq
v=\tilde{f} [x+c(v) t] \,\, \qquad 
\eeq
is found, where
\beq
\qquad c(v)=  \frac{(c_1^2 c_4+c_2c_3^2)\sigma^2}{ c_1 (c_1 v+c_3 \sigma)^2}    - \frac{c_2 }{c_1}\,\,  
\eeq
is a rational characteristic speed and $\tilde{f} $ is an arbitrary function of its argument  that is typically provided in the form of initial datum. Remarkably, if the quantity
\beq
\Delta=c_1^2 c_4+c_2 c_3^2  
\label{def Delta}
\eeq
vanishes one merely has that the initial datum $\tilde{f}(x)$ propagates at constant speed, 
i.e. $v=f\left(x-\frac{c_2}{c_1}t\right)$. When instead $\Delta\neq 0$ 
then $\frac{d}{dv}c(v)\neq0 $, and compressive shock or espansive fan solutions are in principle contemplated \cite{whitham}. 
 
\section{Similarity reductions }
\label{section similarity reductions}

In this Section, we would like to pay attention on a special subclass of solutions to Equation \eqref{veq}:  those that are connected with the reduction of the equation on the orbits of its group symmetry generators. The knowledge of symmetry generators underlying a given differential problem generally proves to be helpful in two complementing respects, indeed. The first possibility is  to construct new solutions from known  ones \cite{vinogradov, olver}.  Besides, one can perform a {\em similarity reduction} of the differential problem  to find solutions that remain invariant under the action of the symmetry group, known as \emph{similarity solutions}. The identification of such reductions relies on the recogniton of the class of invariants $I(x,t,v)$ to a given symmetry generator $W$, that is  such that  $W I(x,t,v)=0$ \cite{olver}. Below, we will proceed in this second respect.

\indent 
Symmetry generators for Equation \eqref{veq} with $c_1c_2c_3c_4\neq0$ have been computed in \cite{giglio jpa}. It turns out that  
a 3-parameter group of symmetry underlies the differential problem (\ref{veq})  through the action of the following three  vector fields: 
\beq
W_1= \partial_x \,\, , \qquad  W_2= \partial_t + \frac{c_2}{c_1} \partial_x \,\,, \qquad 
W_3= t \partial_t +  \left( 2\frac{c_2}{c_1} t -  x \right) \partial_x +  \, \left(\frac{c_3 \sigma}{c_1} +v\right)\partial_v \,\,.
\label{W1 W2 W3}
\eeq
Each of these operator defines a one-dimensional subgroup $G_{k}(\{x,t,v\}; \lambda)$ of  local  transformations depending on a single real parameter 
$\lambda_k$ ($k=1,2,3$).\footnote{When $\Delta= c_1^2 c_4+c_2 c_3^2 =0$,  infinite symmetries come into play: in addition to three generators 
(\ref{W1 W2 W3}), the family of symmetry generators  
 \beq 
W^\infty= \left[ G_1+\frac{c_1}{c_2} F_1(t)\right] \,  \frac{\partial}{\partial t}+\left[   F_2\left(-\frac{c_1}{c_2} x+t \right)
   +F_1(t) \right] \, \frac{\partial}{\partial x} +
   \frac{c_1}{c_2} \left( v\;+\;\sigma \frac{c_3}{c_1} \right) F_2' \left(-\frac{c_1}{c_2} x+t\right)\, 
   \frac{\partial}{\partial _{v} } 
   \label{W infty}
   \eeq
is found, being    $F_1, \;F_2$  arbitrary functions of their argument and  $G_1$ constant.\label{footnote_Winfty}
} 

\subsection{Similarity reductions from $W_1$ and $W_2$}
\label{subsection stationary limit}
Operators $W_1$ and $W_2$ are evidently associated with rigid translations in the $x$ and $t-\frac{c_1}{c_2} x$ directions.  
The two corresponding invariants  $I_1=I_1(t,v)$ and  $I_2=I_2(c_1 x-c_2t,v)$ 
point to the  natural one-dimensional reductions $v=v(t)$  and $v=v(c_1 x-c_2t )$. However, both reductions yield to solutions of \eqref{veq} that are merely constants. Linear combinations of $W_1$ and $W_2$ can be considered too. In particular, the operator $\partial_t$ can be obtained, whose invariants imply the reduction $v=v(x)$. The singular behaviour of the resulting similarity solutions can be immediately inferred because Equation (\ref{veq}) reduces  to the nonlinear ordinary differential equation of the Riccati type 
\beq
\label{veq2 x}
c_{2} v^{2}  + c_4 \sigma^2+ \eta \sigma   c_{2} \partial_x v    = C_0 \, ( c_{1} v + c_{3}  \sigma )\, , \, 
\eeq
where $C_0$ is a constant. By virtue of this, when $C_0 c_2\neq0$ solutions $v(x)$  to equation (\ref{veq2 x}) are  
\beq
v(x)=\frac{ c_1 C_0}{2c_2} -\frac{\tilde{C}_0}{2c_2}  \tan \left[ \frac{c_1 \tilde{C}_0}{2} 
\left(\frac{x}{c_2 \eta  \sigma }+C_1 \right)\right]\,\,, 
\label{sol tan x}
\eeq
where $C_1$ is an integration constant and $ \tilde{C}_0=\sqrt{4 \sigma c_2 (\sigma c_4 -C_0 c_3)-C_0^2c_1^2}\neq0 $. 
If, in contrast, $\tilde{C}_0=0$, i.e. if $C_0$ takes one of the values
$$
C_{0}^{(\pm)}=\frac{2}{c_1^2}  \left[-c_2 c_3 \sigma \pm \sqrt{c_2 \left(c_1^2 c_4+c_2 c_3^2\right) \sigma ^2}\right] \,\, , 
$$
then one has the simple rational structure
\beq
v_\pm=  \,\frac{c_1 \, C_{0}^{(\pm)}}{2 \,c_2}+\frac{\eta \sigma }{ x- C_1} \,\, . 
\label{sol stazionarie pm}
\eeq 
The  existence of singular real solutions to \eqref{veq} is thus put in evidence immediately through \eqref{sol tan x} and \eqref{sol stazionarie pm} for real 
$C_{0}$, $C_1$, $\tilde{C}_0$ and $C_\pm$.
Analogous results follow from the reductions implied by the invariants associated with other linear combinations of $W_1$ and $W_2$.
\subsection{Similarity reductions from $W_3$}

The explicit one-parameter group of symmetry transformation implied by the generator $W_3$ is also straightforwardly inferred, 
\beq
G_3(\{x,t,v\};\lambda_3)=\left\{ x_{\lambda_3}=e^{-\frac{c_1}{c_2}\lambda_3} x +2\, t\, \frac{c_2}{c_1} \, 
\sinh\left(\frac{c_1}{c_2}\lambda_3\right)  \,, \,\, t_{\lambda_3}=e^{\frac{c_1}{c_2}\lambda_3} t \,, \,\,v_{\lambda_3}
= \,\,e^{\frac{c_1}{c_2}\lambda_3} v+\frac{c_3 \sigma}{c_1} (e^{\frac{c_1}{c_2}\lambda_3}-1) \right\}\,\,.
 \label{simmetria G3}
\eeq
But once attention is paid to the similarity solutions for the equation \eqref{veq} that would result from the generator $W_3$,  implications are less expeditious. Indeed, the 
 concerned  invariants  take the form 
\begin{equation}
 I_3=I_3\left[ \left(1+\frac{c_1}{c_3 \sigma} v\right) \,\,t^{-1}\, ,\,\frac{\eta}{c_3} t (c_1 x-c_2 t)\right] \,\, . 
\label{I2}
\end{equation}
According to (\ref{I2}),  and by a convenient normalisation, functions $v$  can  be sought  of the form
\beq
v=\frac{ \sigma c_3}{c_1}  \left[\, B \, \frac{\eta c_1^2}{c_3} \, t \, \Phi (\xi)-1\right]\,\,\ , \qquad \xi=B   \, t \,(c_1 x-c_2 t) \,\, 
\label{def xi}
\eeq
with $B$ real constant. By doing so, Equation (\ref{veq})  turns into the following  second order nonlinear ODE for $\Phi$ 
\beq
\Phi ''(\xi )=\frac{\Phi '(\xi )^2}{\Phi (\xi )} -\left[\Phi (\xi )+\frac{1}{\xi }-
\frac{  \Delta  }{B \, c_1^4 \, \eta^2   }\, \frac{  1}{  \xi \, \Phi (\xi )}\right]\, \Phi '(\xi ) -\frac{\Phi (\xi )^2}{\xi }
\qquad \qquad \left('=\frac{d}{d\xi} \right) \,\, , 
\label{new EQ riduzione xi}
\eeq
resembling a Painlev\'e III (PIII) equation with all coefficients trivial but one  (see e.g. \cite{conte2020,clarkson, huard}). 
We are thus lead to distinguish two cases on the basis of the values attained by coefficients. Depending on whether or not   $\Delta=c_1^2 c_4+c_2 c_3^2$ vanishes, two distinct differential equations therefore arise: 
\beit
\item Equation 1: 
\beq
\underbrace{\Phi ''(\xi ) =-\frac{\Phi '(\xi )}{\xi }+\frac{\Phi '(\xi )^2}{\Phi (\xi )}-\frac{\Phi (\xi )^2}{\xi } }_{\text{PIII}}  -\Phi (\xi ) \Phi '(\xi )
\label{new EQ riduzione xi 1}
\eeq
for $\Delta=0$ whatever the constant $B$ in \eqref{new EQ riduzione xi}; 
\item  Equation 2: 
\beq
\underbrace{
\underbrace{\Phi ''(\xi )=-\frac{\Phi '(\xi )}{\xi }+\frac{\Phi '(\xi )^2}{\Phi (\xi )}-
\frac{\Phi (\xi )^2}{\xi }}_{\text{PIII}} -\Phi (\xi ) \Phi '(\xi ) }_{\text{Equation \eqref{new EQ riduzione xi 1}}}
+\frac{\Phi '(\xi )}{\xi  \Phi (\xi )}
\label{new EQ riduzione xi 2}
\eeq
when $\Delta\neq0$,  upon setting for convenience 
\beq
B=\frac{  \Delta }{ \eta^2 c_1^4} \,\,.
\label{B caso 2}
\eeq 
\enit
In both \eqref{new EQ riduzione xi 1}-\eqref{new EQ riduzione xi 2} the similarity with the Painlev\'e III  equation $\text{P}_{\text{III}}(\xi; -1,0,0,0)$  with all but one  vanishing coefficients is underlined. It is noteworthy that the constraint $\Delta=0$ has already proved to be relevant in the study of  \eqref{veq} as it gives rise to infinite local symmetries connected to the equation  \cite{giglio} (see Eq. \eqref{W infty} in footnote \ref{footnote_Winfty}).
 
 In next sections we shall tackle the problem of understanding properties of solutions to \eqref{new EQ riduzione xi 1} and \eqref{new EQ riduzione xi 2}, being heedful to the occurrence of singularities.

\section{Properties of Equation (\ref{new EQ riduzione xi 1}) and of its solutions}
  \label{section Delta 0}
While Equation~(\ref{new EQ riduzione xi 1})  may appear at first to be uneasy to solve, 
its solutions  prove to possess a simple analytical form that can be straightforwardly determined. 
The general real solution reads indeed:
\beq
\Phi_0(\xi; \alpha_1,\alpha_2)=\frac{\alpha_1-1}{\alpha_2(\alpha_1-1)  \xi ^{\alpha_1}-\xi } \,\,\ , \qquad \xi=   \,  t \,(c_1 x-c_2 t)
\label{sol Phi caso Delta 0}
\eeq
where $\alpha_{1}$ and $\alpha_2$ are real-valued constants of integration  and the arbitrary constant $B$  introduced in 
 \eqref{def xi}  has been normalised to unity for simplicity.  Notice that the 2-parameter family of real functions \eqref{sol Phi caso Delta 0} comprises rational functions 
with simple single poles, namely $\Phi_0=(\alpha_1-1)\xi^{-1}$ or  $\Phi_0=(\xi+\alpha_2)^{-1}$, which can be obtained for $\alpha_2=0$ and $\alpha_1=0$, respectively. The null solution function arises instead for $\alpha_1=1$. It is also worth to point out that a deeper connection between \eqref{new EQ riduzione xi 1} and the Painlev\'e III equation unfolds through a Cole-Hopf transformation  $\Phi(\xi)=[ \log (A-W(\xi))]'$ (with $A$ arbitrary constant): the derivative $w(\xi)=W'(\xi)$ of the Cole-Hopf potential function $W(\xi)$ obeys indeed $w''=(w')^2 w^{-1}-w' \xi^{-1}$,  the Painlev\'e III equation with all parameters null.

Before to proceed in shedding light on features of functions \eqref{sol Phi caso Delta 0},
the remark is in order  that rational non integer values of $\alpha_1$ make $\Phi_0$ multivalued \cite{morse}. Recalling that \eqref{sol Phi caso Delta 0} is the solution to a nonlinear ODE (Equation~\eqref{new EQ riduzione xi 1}) with prescribed initial conditions that may possibly arise from  specific applicative models, we will  simply restrict our study to $\xi \in \mathbb{R}$  and $\Phi (\xi) \in \mathbb{R}$, without providing a criterion for selecting specific real roots. In fact, when multi-valuedness occurs, specific representations would be  naturally gauged from the  motivating application. 

Quantitative and qualitative properties of functions \eqref{sol Phi caso Delta 0} are discussed separately in the following two Subsections. 

\subsection{Quantitative properties of solutions  \eqref{sol Phi caso Delta 0}}
\label{subsection quantitative}

We proceed by analysing the solution~\eqref{sol Phi caso Delta 0} for positive real $\alpha_2$ and rational $\alpha_1$ evaluating singularities and stationary points as they vary, starting with the simplest case $\alpha_1\in \mathbb{Z}$.  For simplicity, we will consider  the case $\alpha_2>0$. A similar scenario, which is not discussed in detail in this work,  is expected when $\alpha_2<0$. Singularities of solution~\eqref{sol Phi caso Delta 0} for  integers $\alpha_1$ are given by the following, accounting for the standard classification of singularities in the complex plane \cite{kevorkian,morse}.

\begin{proposition}[Singularities of solution~\eqref{sol Phi caso Delta 0} for integer values of  $\alpha_1$]
	\label{poles}
	Let $\alpha_1 \in \mathbb{Z}$, $\alpha_2>0$, $\xi^{(0)} =0$ and  $\xi_{\pm} =\pm | (1-\alpha_1)  \alpha_2|^{\frac{1}{1-\alpha_1 }}$.  The real singularities of the function \eqref{sol Phi caso Delta 0}   are listed below.
	
	\begin{enumerate}[i)]
		\item  No singularities on the real line for odd negative integer $\alpha_1$.
		\item One single singularity (pole) for $\alpha_1$ even negative integer  located at $\xi=\xi_{-}$.  
			    \item Two real singularities when $\alpha_1$ is even positive integer. These are  located at $\xi=\xi^{(0)}$  (pole) and $\xi=\xi_+$ (branch point).
		\item Three singularities on the real line for odd positive integers $\alpha_1$. One is located at $\xi=\xi^{(0)}$ (pole) and the other two at $\xi=\xi_{\pm}$ (branch points).
	    \end{enumerate}
\indent 
 \begin{proof}
    Singularities of \eqref{sol Phi caso Delta 0} for negative integer values of $\alpha_1$ are identified by  real roots of the polynomial $ \mathcal{P}^{-} (\xi):=  \xi ^{1-\alpha_1}-\alpha_2(\alpha_1-1)$, which has complex roots $\xi_l=|  (1-\alpha_1) \alpha_2 |^{\frac{1}{1-\alpha_1}} e^{i\frac{\pi+ 2 l \pi}{1- \alpha_1}}$, $l=0, 1, \cdots, -\alpha_1$. 
    	 If   $\alpha_1$ is odd, there are no real roots\footnote{An equivalent way to see this consists in observing that $\alpha_2(\alpha_1-1)< 0$. If   $\xi \in\mathbb{R}$ and  $\alpha_1$ is odd, then  $1-\alpha_1$ is even and  $\underbrace{\xi ^{1-\alpha_1}}_{>0}=\underbrace{\alpha_2(\alpha_1-1)}_{<0}$ is inconsistent.}, hence proving i). If $\alpha_1$ is even instead, a real solution is obtained when $\pi+ 2 l \pi=(1- \alpha_1)\pi$, that is for $l=- \frac{\alpha_1}{2}$, giving $\xi_{-} =- | \alpha_2 (1-\alpha_1) |^{\frac{1}{1-\alpha_1}}$. This proves case ii). \\
    	 Similarly to cases i) and ii), real poles of \eqref{sol Phi caso Delta 0} for positive integer values of $\alpha_1$ are given  by  real roots of the polynomial $ \mathcal{P}^{+} (\xi):=  \xi \left[ \left( (\alpha_1 -1)  \alpha_2 \right) \xi^{\alpha_1-1}-1 \right] $. Excluding the trivial case $\alpha_1=1$, which corresponds to the null solution $\Phi_0(\xi)=0$, the root $\xi^{(0)}=0$ is readily identified for all values of $\alpha_1$. The other complex roots of $ \mathcal{P}^{+} (\xi)$ are $\xi_l=| (1-\alpha_1) \alpha_2 |^{\frac{1}{1-\alpha_1}} e^{i\frac{ 2 l \pi}{1- \alpha_1}}$, $l=0, 1, \cdots, \alpha_1-2$. A real root is promptly obtained for $l=0$, that is  $\xi_{+} = |  (1-\alpha_1)\alpha_2 |^{\frac{1}{1-\alpha_1}}$. Another  solution  for odd $\alpha_1$ is found requiring $2 l \pi=(1-\alpha_1)\pi$. Such solution   reads $\xi_{-} =- | (1-\alpha_1) \alpha_2 |^{\frac{1}{1-\alpha_1}}$, hence completing the proof of iii) and iv).  
    		\end{proof}
		\end{proposition}

Singularities $\xi_\pm$ are clearly {\it movable} in that they are not fixed by the equation \eqref{new EQ riduzione xi 1},  but rather they are determined by the initial condition assigned for the equation itself. In fact, the only {\it essential} singularity, when present, is at $\xi^{(0)}=0$, as one can see from  \eqref{new EQ riduzione xi 1}.   

 Properties of  solution \eqref{sol Phi caso Delta 0} can be further characterised by looking at its stationary points. The stationary points of solution \eqref{sol Phi caso Delta 0} for integer $\alpha_1$ are provided by the following:

\begin{proposition}[Stationary points  of solution~\eqref{sol Phi caso Delta 0} for integer values of $\alpha_1$]
	\label{stationary points}
	Let $\alpha_1 \in \mathbb{Z}$, $\alpha_2>0$, $\xi^{(0)} =0$ and  $\xi_{\pm}^c =\pm |  \alpha_1 (1-\alpha_1)\alpha_2|^{\frac{1}{1-\alpha_1 }}$.  The stationary points of solution \eqref{sol Phi caso Delta 0}   are listed below.
	
	\begin{enumerate}[i)]
		\item  Three stationary points located at $\xi=\xi^{(0)}$ (inflection point) and  $\xi=\xi_{\pm}^c$ for $\alpha_1$ odd  negative integer with $\alpha_1<-1$.
				\item  Two stationary points when $\alpha_1=-1$, located at $\xi=\xi_{\pm}^c$ .
		\item  Two stationary points for $\alpha_1$ even negative integer,  located at $\xi=\xi^{(0)}$ and $\xi=\xi_{+}^c$.  
		\item One single stationary point located at $\xi=\xi_+^c$  for $\alpha_1$ even positive integer.
		\item Two stationary points  for odd positive integers $\alpha_1$ located  at $\xi=\xi_{\pm}^c$.
	\end{enumerate}
\indent
 
\end{proposition}

\noindent Singularities and stationary points of \eqref{sol Phi caso Delta 0} for rational values of $\alpha_1$ can be also derived 
as shown in the below: 

\begin{proposition}[Singularities of \eqref{sol Phi caso Delta 0} for rational values of $\alpha_1$] Let $\alpha_1=\frac{q}{p}$ with $q, p \in \mathbb{Z}$ and $\gcd(q,p)=1$, and consider  $\xi^{(0)}$, $\xi_{\pm}$ defined as in Proposition~\ref{poles}. The real  poles of  \eqref{sol Phi caso Delta 0} are listed below.
	\label{poles rational} 	
	\begin{enumerate}[i)]
				\item Case $\alpha_1<0$.
		\begin{enumerate}[a)]
			\item One singularity located  at $\xi=\xi_-$  (branch point) if $p$ is odd and $q$ is even.
			\item No singularities if $p$ and  $q$ are odd.
			\item No singularities if $p$ is even\footnote{As we have already pointed out, in this work we are standardly considering for $\xi\geq0$ and $p$ even, $\xi^{1/p}=\sqrt[p]{\xi}$. If  $\xi^{1/p}=-\sqrt[p]{\xi}$ is adopted instead, a second singularity arises at  $\xi=\xi_+$.}.
		\end{enumerate}
		\item 
		Case $0<\alpha_1<1$.  
		\begin{enumerate}[a)]
			\item Two singularities if $p$ is odd and $q$ is even located  at $\xi=\xi^{(0)}$ (branch point) and  $\xi=\xi_{-}$ (branch point).
			\item One singularity   located  at $\xi=\xi^{(0)}$ (branch point) if $p$ and  $q$ are odd.
			\item One singularity located at $\xi=\xi^{(0)}$ (branch point) if $p$ is even\footnote{If  $\xi^{1/p}=-\sqrt[p]{\xi}$ is adopted, a second singularity arises at  $\xi=\xi_+$.}.
		\end{enumerate}
 
	\item Case $\alpha_1>1$.
\begin{enumerate}[a)]
	\item Two singularities if $p$ is odd and $q$ is even located  at $\xi=\xi^{(0)}$ (pole) and  $\xi=\xi_{+}$  (branch point).
	\item Three singularities if $p$ and  $q$ are odd  located  at $\xi=\xi^{(0)}$ (pole) and $\xi=\xi_{\pm}$ (branch points).
	\item Two singularities  at $\xi=\xi^{(0)}$ (pole) and $\xi=\xi_+$ if $p$ (branch point) is even.
\end{enumerate}
		\end{enumerate}
 
	\end{proposition}

\begin{proposition}[Stationary points of \eqref{sol Phi caso Delta 0} for rational values of $\alpha_1$] 
Let $\alpha_1=\frac{q}{p}$ with $q, p \in \mathbb{Z}$ and $\gcd(q,p)=1$, and consider  $\xi^{(0)}$, $\xi_{\pm}^c$ defined as in Proposition \ref{stationary points}. The stationary points of  \eqref{sol Phi caso Delta 0} are listed below.
	\label{stationary points rational} 	
	\begin{enumerate}[i)]
		\item Case $\alpha_1<0$.
		\begin{enumerate}[a)]
			\item Two stationary points located  at $\xi=\xi^{(0)}$ and  $\xi=\xi_+^{c}$  if $p$ is odd and $q$ is even.
			\item Three stationary points located at $\xi=\xi^{(0)}$ (inflection point) and  $\xi=\xi_{\pm}^{c}$  if $p$ and  $q$ are odd.
			\item Two stationary points located  at $\xi=\xi^{(0)}$ and  $\xi=\xi_+^{c}$  if $p$ is even.  
			 		\end{enumerate}
		\item 
		Case $0<\alpha_1<1$.  
		\begin{enumerate}[a)]
			\item One stationary point at $\xi=\xi_-^{c}$ if $p$ is odd and $q$ is even.
			\item No stationary points  if $p$ and  $q$ are odd.
			\item No stationary points  if $p$ is even\footnote{If one considers $\xi^p=-\sqrt[p]{\xi}$ a stationary point is located at $\xi=\xi_+^{c}$.}.
		\end{enumerate}
		 
		\item Case $\alpha_1>1$.
		\begin{enumerate}[a)]
			\item One stationary point located at  $\xi=\xi_+^{c}$  if $p$ is odd and $q$ is even.
			\item Two stationary points   located  at $\xi=\xi_{\pm}^c$  if $p$ and  $q$ are odd.
			\item One stationary point at $\xi=\xi_+^c$  if $p$  is even.
		\end{enumerate}
	\end{enumerate}
	\end{proposition}

Proofs of Propositions \ref{stationary points}-\ref{stationary points rational}   can be given, {\it mutatis mutandis}, 
in a similar fashion of  Proposition \ref{poles}  and are reported in Appendix \ref{appendice 2}.
Notice that  statements in Proposition \ref{poles} for integer values of $\alpha_1$ can be also deduced from Proposition~\ref{poles rational}. For instance, case iii) in Proposition~\ref{poles}, i.e. $\alpha_1$ even positive integer, can be obtained from case  iii.a) requiring $p$ odd (precisely $p=1$) and $q$ even.

Depending on the values of  constants $\alpha_1$ and $\alpha_2$,   distinct features are thus displayed for functions \eqref{sol Phi caso Delta 0}.   
Some symmetry properties are promptly perceived:  an intertwining creates between the problems of revealing asymptotes and stationary points, and mirror situations are designed through sign changes. In particular, remark that one may write
\beq
  \Phi_0'(\xi; \alpha_1,\alpha_2) = -\frac{\Phi_0^2(\xi; \alpha_1,\alpha_2) }{
\xi\Phi_0(\xi; \alpha_1,\alpha_1\alpha_2 ) }\,\, .
\eeq

\subsection{Qualitative properties of solutions \eqref{sol Phi caso Delta 0}}

In this Subsection we provide a more circumstantial picture of solutions \eqref{sol Phi caso Delta 0} by  complementing with figures previous results concerning the individuation of their singularities and stationary points.

\begin{figure}[h!]
\begin{center}
 \includegraphics[scale=0.6]{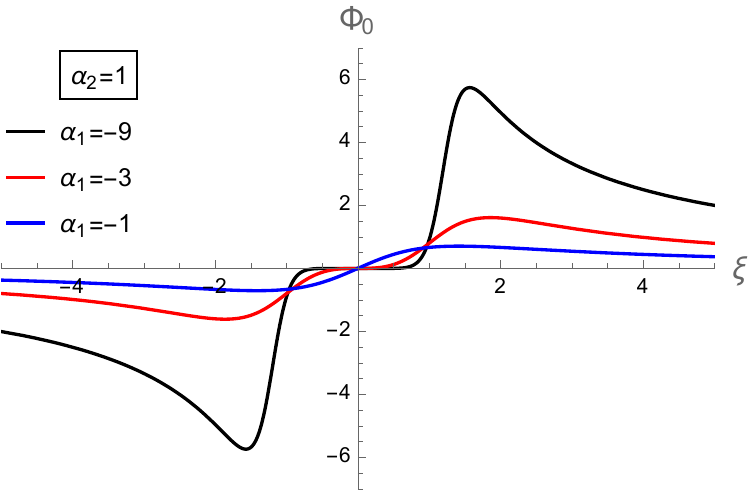}  \,\, a)
  \includegraphics[scale=0.6]{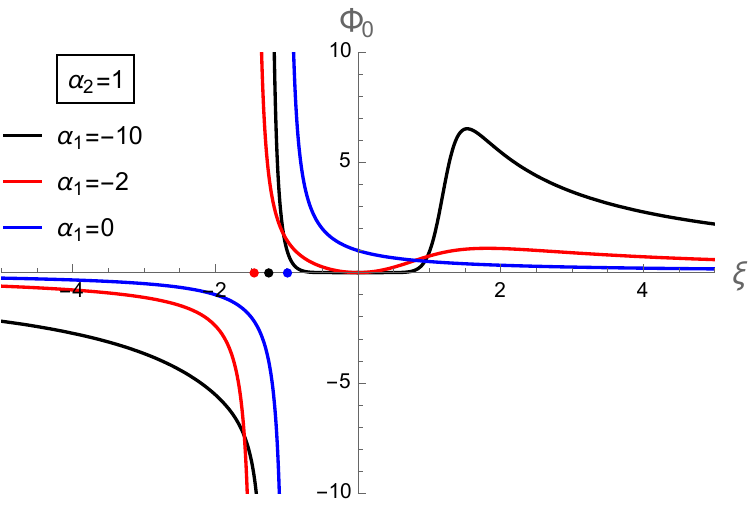} \,\, b)
    \includegraphics[scale=0.6]{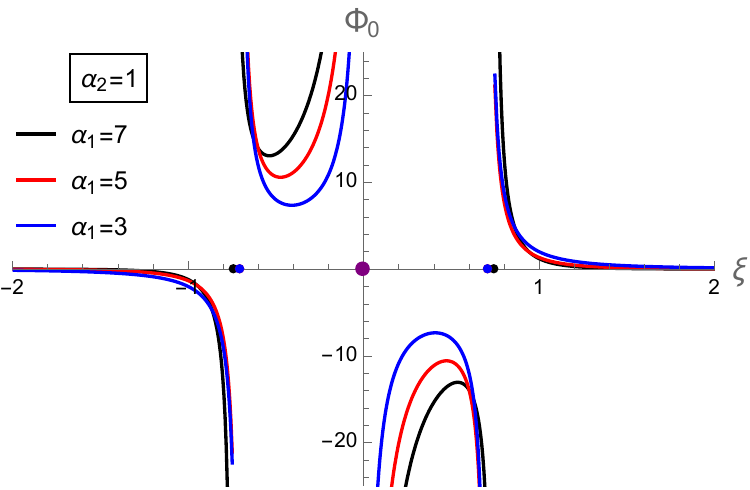}  \,\, c)
  \includegraphics[scale=0.6]{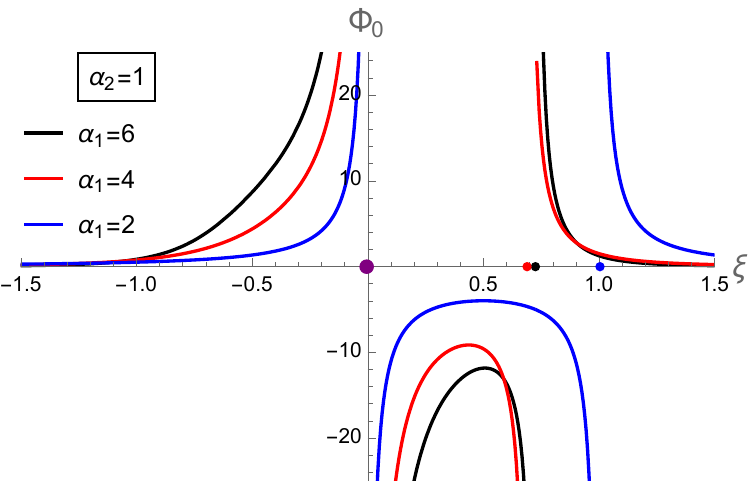} \,\, d)\\
   \includegraphics[scale=0.6]{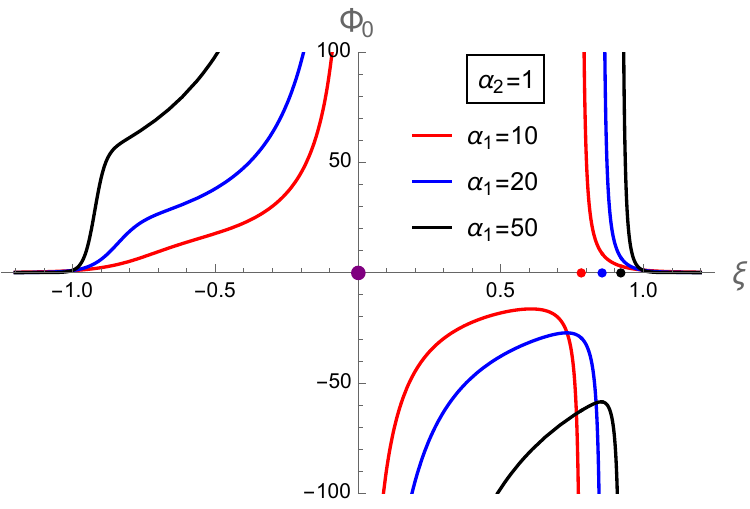}  \,\, e)
   \includegraphics[scale=0.62]{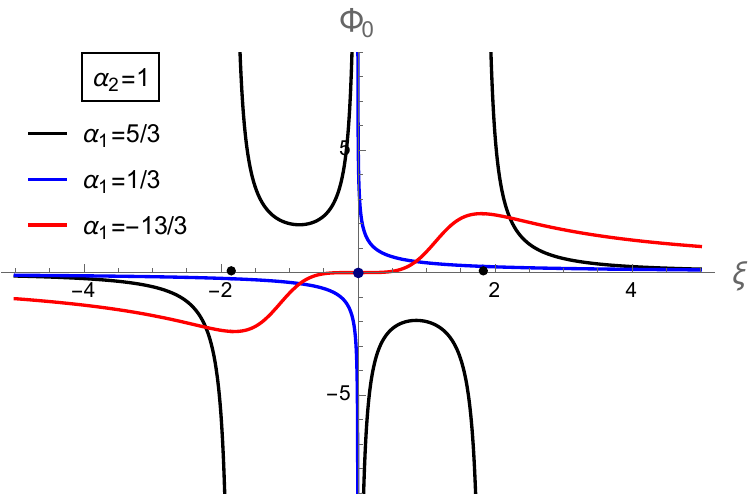}  \, f)
        \includegraphics[scale=0.6]{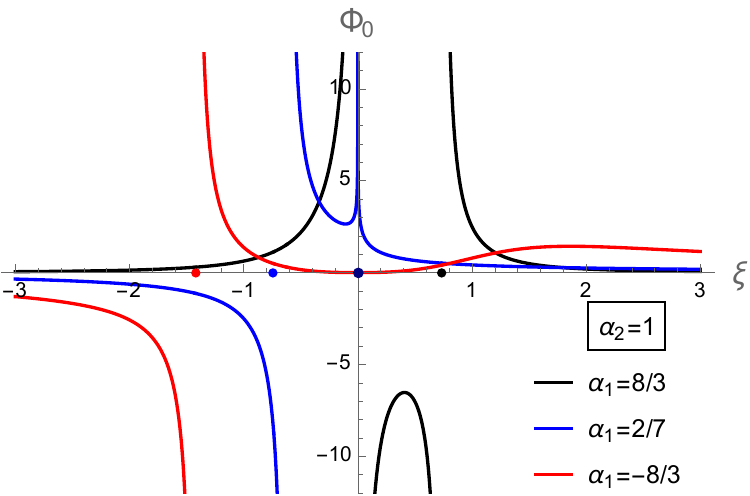}  \,\, g)
   \includegraphics[scale=0.6]{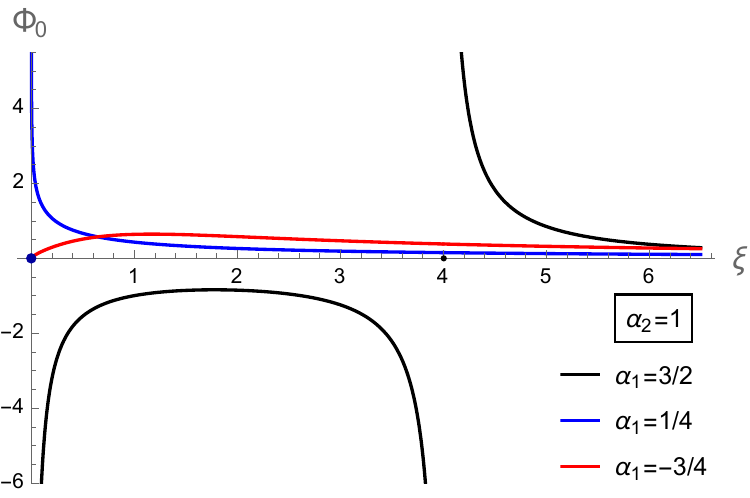}  \,\, h)\\

\caption{ Plots of function $\Phi_0(\xi)$ of Eq. \eqref{sol Phi caso Delta 0} for $\alpha_2=1$.  Colored dots refer to the asymptotes location  evinced through Propositions \ref{poles} and \ref{poles rational} given in Subsection \ref{subsection quantitative}.  a) $\alpha_1$ negative odd integer;
b) $\alpha_1$ negative even integer;
c) $\alpha_1$ positive odd integer;
d) $\alpha_1$ positive even integer; 
e) Birth of a step for higher values of $\alpha_1$ positive even integer;  
  f)-h) Rational values $\alpha_1=q/p$ with $q$ and $p$ integers s.t. $\gcd(q,p)=1$:  f)  $q$ and $p$ both odd; g)  $p$ odd and $q$ even;  h) $p$ even and $q$ odd. 
}
\label{plotPhiDelta01}
\end{center}
\end{figure}

In Figure \ref{plotPhiDelta01} it is shown what is implied for the function $\Phi_0$ of Equation \eqref{sol Phi caso Delta 0} 
whenever $\alpha_2>0$ (which, for simplicity, has been set there to the unit value without loss generality).
  Plots there reveal  up to three vertical asymptotes, and the possibility to develop local maxima and minima, 
 as reckoned in the previous Subsection \ref{subsection quantitative}.  First five plots refer to the case where  $\alpha_1$ is an integer number, 
for which Propositions \ref{poles} and \ref{stationary points} hold. 
 Figure \ref{plotPhiDelta01}.a) displays  the continuous curves generated for negative odd integers 
$\alpha_1$, and  the amplification is marked as long as $\alpha_1$  takes lower and lower values. The effect magnifies itself about the local maximum and minimum points, located at
 $ \xi_\pm^{c} =\pm | \alpha_1 (1-\alpha_1) \alpha_2 | ^{\frac{1}{1-\alpha_1}} $,  showing the transition from smoother curves to shapes with sharp peaks. The changes call for a net 
bending towards the $\xi$-axis of wider portion of curves about the origin before to reach the stationary points at faster rate.  The null asymptotic values $\Phi_0(\xi\to 0; \alpha_1,\alpha_2)=0$ are approached close later of course.  Curves turn out to be symmetric under the combined action of reflections of the dependent and independent variables, $\xi\to -\xi$ and $\Phi_0\to -\Phi_0$.  
Figure \ref{plotPhiDelta01}.b) indicates instead what happens for negative even integers. Comparison is made with the rational function with single pole in $\xi_0=-\alpha_2$ obtained for $\alpha_1=0$. While the right portion of profiles share the features discussed for the prior case, something different happens in the negative $\xi$-domain, where  a vertical asymptote generates and no minimum forms. Decreasing $\xi$ from the origin, the function $\Phi_0$ does no longer decrease but is lifted up, with a final boost on its rate while getting sufficiently close to the vertical asymptote.  
Passing to the positive odd integer $\alpha_1$, see Figure \ref{plotPhiDelta01}.c) , we go back to a picture where $\Phi_0$ is an odd function of its 
argument $\xi$.  Both a maximum and a minimum are comprised, but the curve splits into four portions owing to the appearance of three vertical asymptotes, at the origin and at points 
${\xi}_{\pm}=\pm |  (1-\alpha_1)\alpha_2|^{\frac{1}{1-\alpha_1}}$.   Two vertical asymptotes are rather concerned for positive even integers $\alpha_1$, at the origin and at $\xi^c_+$, a negative maximum laying down in between, see Figure \ref{plotPhiDelta01}.d). Moreover,  while  moving to higher values of $\alpha_1$, a step  progressively forms about $\xi=-1$, as visible in Figure \ref{plotPhiDelta01}.e) and consistently with \eqref{sol Phi caso Delta 0}.  

The remaining three plots, Figs. \ref{plotPhiDelta01}.f)-h),  deal with the case $\alpha_1\in \mathbb{Q}$, thus relying on Propositions \ref{poles rational} and \ref{stationary points rational} in Subsection \ref{subsection quantitative}. To proceed, we assign to  each 
 $\alpha_1\in \mathbb{Q}$ a unique pair of integers $q,p$ such that $\alpha_1=q/p$, with $p>0$ and  $\gcd(q,p)=1$. When   $q$ and $p$ are both  odd  the analogy is with the case of odd integers $\alpha_1$. Precisely, if $q$ is odd positive (see black curve in Fig.  \ref{plotPhiDelta01}.f)) then the behavior resembles the one of of Figure \ref{plotPhiDelta01}.c) for odd integers $\alpha_1>1$. In a similar way, when $q$ is an odd negative (red curve in Fig.  \ref{plotPhiDelta01}.f) the likeness is with \ref{plotPhiDelta01}.a)  treating odd negative integers $\alpha_1$.\footnote{The cases $p=1$ should indeed reproduce what is known for integers $\alpha_1$.}
 Correspondingly, when $p$ is odd and $q$ is even, the behavior of even integers $\alpha_1$ is recovered. For instance, when $\alpha_1<0$ the curves originating for 
$\Phi_0(\xi;\alpha_1,1)$ (red curve in Fig.  \ref{plotPhiDelta01}.g)) are alike those for negative even $\alpha_1$ in Fig. \ref{plotPhiDelta01}.b). On the contrary, 
for positive $\alpha_1<0$ (black curve in  \ref{plotPhiDelta01}.g)) the resemblance is with shapes arising  for even positive $\alpha_1$ (Fig. \ref{plotPhiDelta01}.d)). 
Outcomes when $0<\alpha_1<1$  have to be commented separately. Blue curves in Figs. \ref{plotPhiDelta01}.f)-h) show the singularity at the origin $\xi^{(0)}=0$, and possibly a second one at  $\xi_-^c$ if  $p$ and $q$ are odd and even, respectively. 
In the latter occurrence, the singularity at $\xi_-$  remains visible for small values of $\alpha_1$, in that $\xi_{-}=- |  (\alpha_1-1)\alpha_2|^{\frac{1}{1-\alpha_1}} \simeq - |\alpha_2|$. 
 In contrast, $\xi_- \to  0^-$  when  $\alpha_1$ approaches the unity from below.

\begin{figure}[h]
\begin{center}
 \includegraphics[scale=0.6]{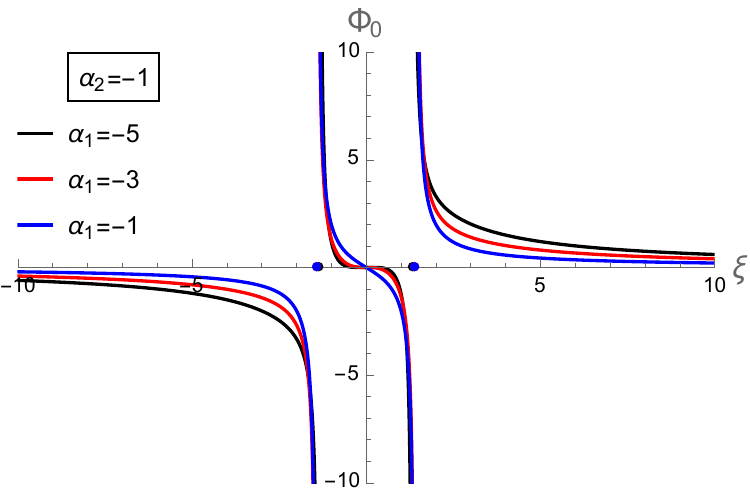}  \,\, a)
    \includegraphics[scale=0.6]{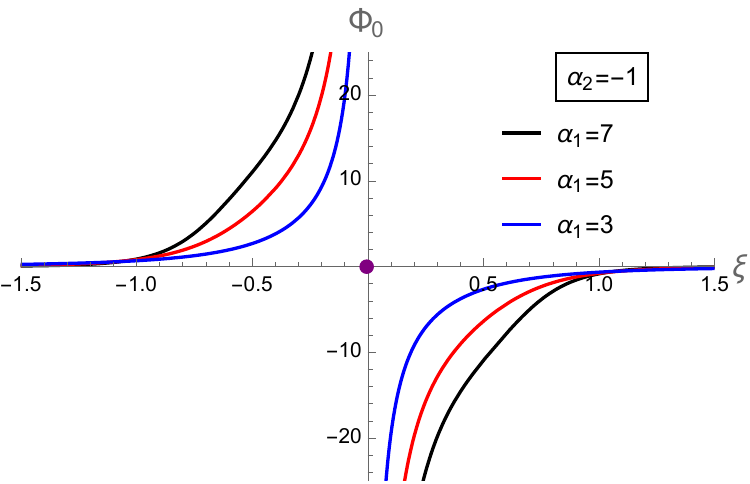}  \,\, b)
\caption{  Plots of function $\Phi_0(\xi)$ of Eq. \eqref{sol Phi caso Delta 0} for $\alpha_2=-1$. Colored dots refer to the asymptotes location. 
a) $\alpha_1$ negative odd integer; b) 
$\alpha_1$ positive odd integer. 
}
\label{plotPhiDelta02}
\end{center}
\end{figure}

The scenario arising for  functions  $\Phi_0(\xi)$ given by Eq. \eqref{sol Phi caso Delta 0} with negative values of $\alpha_2$ can be understood from Figure~\ref{plotPhiDelta02}. 
We proceed concisely on the grounds of symmetries and analitical information gained for $\alpha_2>0$. We start  with by remarking indeed that for $\alpha_2<0$ and integers $j$
 \beq
 \Phi_0(\xi; 2 j, \alpha_2)=-\Phi_{0}(\xi; 2j, -|\alpha_2|) \,\, .
 \eeq
Therefore, curves for $\Phi_0$ when $\alpha_2<0$ and $\alpha_1$ are even integers result from plots b) and d) of Figure  \ref{plotPhiDelta01} upon performing joint reflections $\xi \to -\xi$  and $\Phi_0 \to-\Phi_0$.
As for negative odd integers $\alpha_1$, the two vertical asymptotes in Figure \ref{plotPhiDelta02}.a) are placed at $\xi_{\pm}$. Opposite to Fig.  \ref{plotPhiDelta01}.d) which also furnished us two-asymptote  depiction for the function $\Phi_0$, there is no trait designing a stationary point. Finally for the case $\alpha_2<0$, one asymptote and two monotonically growing curves are typified by positive even integer values of  $\alpha_1$, Figure \ref{plotPhiDelta02}.b).

%
%

\section{Solutions (\ref{def xi}) to Equation (\ref{veq}) when $\Delta=0$}
 \label{section Delta 0 v(x,t)}

The clarification of the behavior allowed for $\Phi_0(\xi)$ put the basis for the understanding of solutions $v(x,t)$ to Equation \eqref{veq} that are or the form \eqref{def xi} and pertinent to the case $\Delta=c_1^2c_4+c_2c_3^2=0$. While conveying particulars of $\Phi_0(\xi; \alpha_1,\alpha_2)$ to solutions of \eqref{veq} via  \eqref{def xi}, values and signs of  constants $c_j$ clearly matter. The role of the evolutionary variable $t$ is also evident in altering the magnitude  and moving the poles of resulting solutions $v(x,t)$. As a matter of fact, the changeover from the similarity coordinate $\xi$ to the original independent variables $x$ and $t$ implies that 
to every point in the $\xi$-axis it is associated a curve $x(t)$ in the $(t,x)$ plane (at $t\neq0$): the straight line $c_1 x= c_2 t$ for the point $\xi=0$,  and the composition 
 \beq
x(t)=\frac{1}{c_1} \left(c_2 t +\frac{\xi}{B t}\right) \,\, 
\label{x(t)}
\eeq
of the same translational motion with an hyperbolic curve for any other point $\xi$. Hence
\begin{enumerate}[i)]
\item  when $\xi/Bc_1$ and $c_2/c_1$ are both strictly positive, then $x(t)$ is strictly positive (negative) over all the positive (negative) $t$-domain; 
\item  if $\xi/Bc_1>0$ and $c_2/c_1<0$, then $x(t)>0$ when either  $ t<-\sqrt{-\xi/Bc_2}$ or $0<t<\sqrt{-\xi/B c_2}$; 
\item if $\xi/Bc_1<0$ and $c_2/c_1>0$, then $x(t)>0$ when either $-\sqrt{-\xi/Bc_2}<t<0$ or $t>\sqrt{-\xi/B c_2}$; 
\item  if $\xi/c_1$ and $c_2/c_1$ are both strictly negative, then $x_s$ is strictly negative (positive) over all the positive (negative) $t$-domain;
\item if $\xi=0$, then ${\rm sign} (x(t)) ={\rm sign} (c_1 c_2  t)$.
\end{enumerate} 

\begin{figure}[h]
\begin{center}
 \includegraphics[scale=0.53]{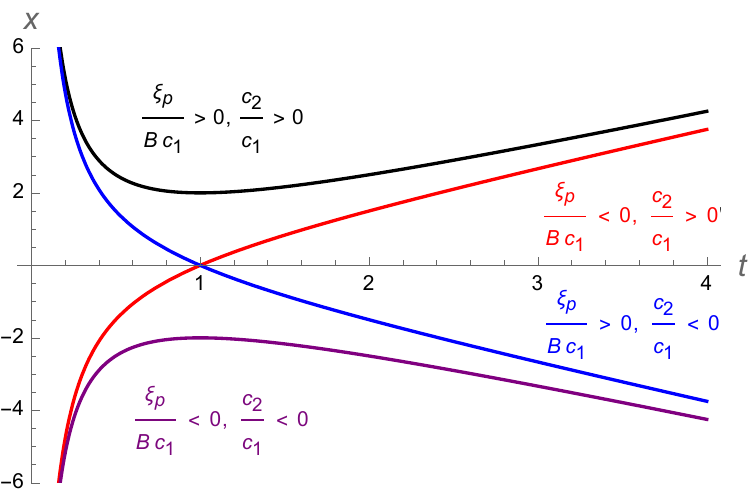}  \,\, a
\caption{Qualitative motion for $ t>0$ of a point $\xi_p$ on the $\xi$-axis  as it is implied by  Equation \eqref{x(t)}.  Case \textit{i)} (black):  $\xi_p/Bc_1 >0$ and $c_2/c_1>0$; 
Case \textit{ii)} (blue):  $\xi_p/Bc_1 >0$ and $c_2/c_1<0$;  Case \textit{iii)} (red):  $\xi_p/Bc_1 <0$ and $c_2/c_1>0$; 
Case \textit{iv)} (purple):  $\xi_p/Bc_1 <0$ and $c_2/c_1<0$. }
\label{moto poli}
\end{center}
\end{figure} 
Qualitative behavior of $x(t)$ concerned with $\xi\neq0$ is summarized in Figure \ref{moto poli}. In particular, the above schematics apply to connote the relocation on the $x$-axis of  singularities  of similarity solutions \eqref{def xi} while $t$ varies. 
We thus see that in cases i) and iv) the same value of $x$ can be obtained for two distinct values of the local variable $t>0$, meaning  that a reversal of the singularity  motion is exhibited.  In cases ii) and iii), instead,  there is a one-to-one correspondence   based on a monotonic behavior for the function $x(t)$ of Equation \eqref{x(t)}.
 

\begin{figure}[h]
\begin{center}
 \includegraphics[scale=0.6]{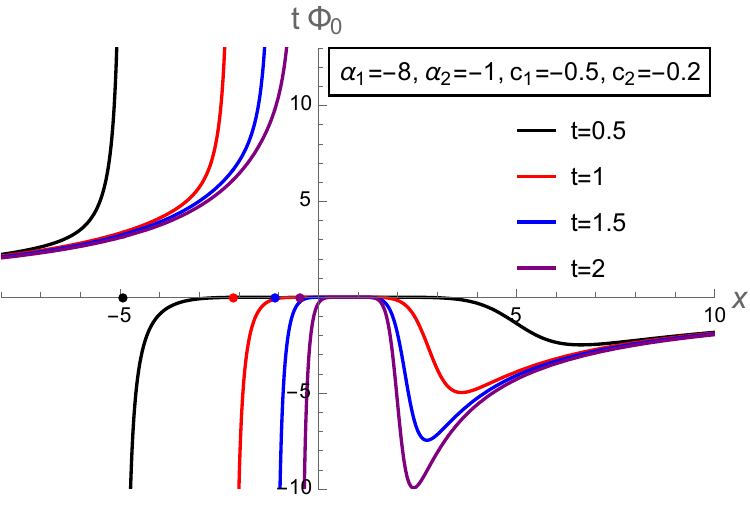}  \,\,  a) 
  \includegraphics[scale=0.6]{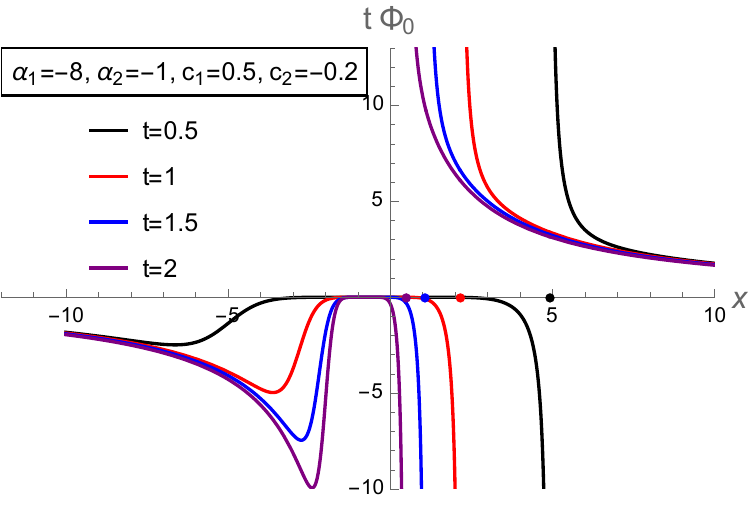}  \,\, b)
 \caption{ Examples of graphs of function $t\Phi(x,t)$ that follows from \eqref{sol Phi caso Delta 0} at different times and for different choices of parameters $c_{1}$ and $c_2$. 
 Equation \eqref{sol Phi caso Delta 0} has been taken with $\alpha_1=-8$ and $\alpha_2=1$,  for which the shape of $\Phi(\xi)$ is analogous to curves of Fig. \ref{plotPhiDelta01}.b). 
 Colored dots refer to the asymptotes location.
 a)  Progressive deformation of shape for $t\Phi(x,t)$  when $c_1$ and $c_2$ are both negative.  b)  Progressive deformation of shape for $t\Phi(x,t)$ when $c_1>0$ and $c_2<0$. 
}
 \label{plottPhiDelta0t}
\end{center}
\end{figure}

\begin{figure}[h]
\begin{center}
 \includegraphics[scale=0.6]{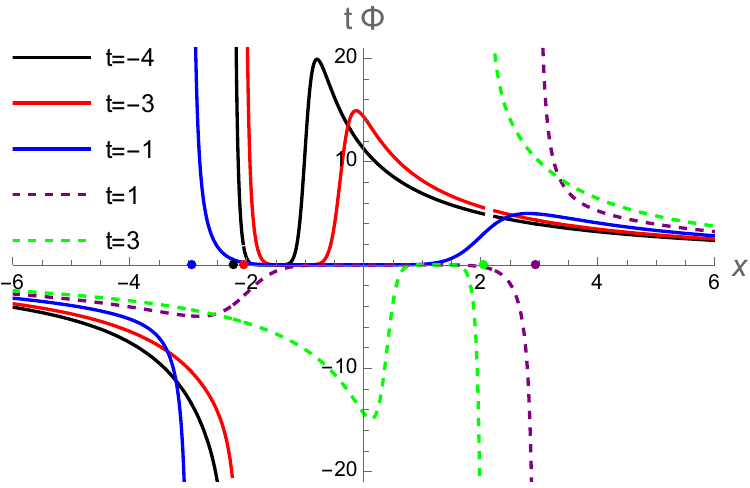}  \,\, a)\quad
 \includegraphics[scale=0.6]{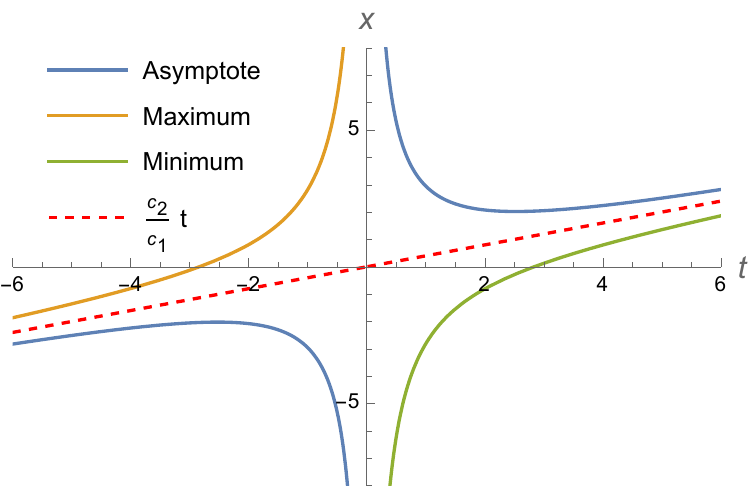}  \,\, b)
 \caption{ Asymptotes and stationary points motion for the function $t\Phi(x,t)$ arising from \eqref{sol Phi caso Delta 0} 
 with $\alpha_1=-8$, $\alpha_2=1$, $c_1=0.5$ and $c_2=0.2$. a) The function $t\Phi_0$ for different values of $t$. For increasing negative $t$ the maximum flattens and shifts the location to higher values of the coordinate $x$. The asymptote moves as well, but 
 after moving to the right on the $x$-axis (see the red curve determined at $t=-3$ on the right of the black one obtained for $t=-4$) it reverses its motion and directs itself to the opposite direction (see the blue curve at $t=-1$ on the left of the previously mentioned black and red ones). b) Motion in the $(t,x)$ plane of asymptote and stationary points, signalling the inversion of motion only for the former.
  }
 \label{pointsmotion}
\end{center}
\end{figure}

We are now in the position to outline more efficiently the significant aspects of similarity solution  \eqref{def xi} when $\Delta=0$. 
Regardless obvious remarks in respect of consequences due to signs of the constants $c_j$,  it is helpful to consider an example of just the function $t\Phi_0=t\Phi_0(x,t)$. In Figure \ref{plottPhiDelta0t}, a case based on curves $\Phi$ of the type Fig. \ref{plotPhiDelta01}.b) is considered, in particular by setting $\alpha_1=-8$ and $\alpha_2=1$. Plots there supply a glimpse into effects of sign changes for the constants $c_{1,2}$ at positive $t$. For completeness, and being more instructive, the spectrum of the real ``evolutionary'' variable $t$ shall be permitted to comprise negative values. To this, we will consider coefficients $c_1$ and $c_2$ be both positive, as shown in Figure \ref{pointsmotion}. 
In the formal limit $t\to -\infty$, resulting function $t\Phi_0(x,t)$ is null. For negative but increasing $t$'s,  profiles  like   curves in Figure \ref{plotPhiDelta01}.b)  tend to form, with lower and lower values attained by the maximum  which moves to the right becoming smoother.  The generated asymptote first moves to the right either, then reverses its motion by progressively placing itself at lower and lower values of $x$. At $t=0$, the asymptote and the maximum (experiencing the ongoing magnitude suppression) are pushed at infinity, and the function lies again on the $x$ axis.  At later $t$, curves are no longer similar to those of Fig. \ref{plotPhiDelta01}.b), but rather to those ones one gets from them through the simultaneous reflections  about horizontal and vertical axes. An asymptote is brought back in the picture from $\infty$, approaching a stationary point which is reinstated, but as as a minimum moving from $-\infty$ towards the direction of the increasing $x$ becoming narrower and narrower.  The asymptote again experiences a ``bouncing back'' mechanism and an inversion of its motion, but without starting to departing from the minimum (see Fig. \ref{pointsmotion}).
In the formal asymptotic limit $t\to \infty$, the curve again flattens down on the abscissa.
 Analogous analysis can be worked out by chosing other pairs of parameters $\alpha_{1}$ and $\alpha_2$. 

\section{Properties of Equation (\ref{new EQ riduzione xi 2}) and  its solutions}
\label{section Delta gen}

Evidently, Equation (\ref{new EQ riduzione xi 2}) is a non-trivial nonlinear  ODE. Contrary to Equation \eqref{new EQ riduzione xi 2}, we have not found a  transformation connecting it to the list of Painlev\'e equations.   Of course approximate methods can be resorted in the attempt to have a glimpse of possible dynamics emerging from it. For instance, Painlev\'e-like  test arguments (see e.g. \cite{conte,hone} and Refs. therein for an account on this subject) can be carried out while aimed at highlighting features of its solutions, such as their singularity structure.  {\em A priori} Equation  (\ref{new EQ riduzione xi 2})   can have  solutions with movable singularities which vary with the initial conditions. Testing the singularity structure of the equation appears particularly meaningful, as it is natural to wonder about 
admissible deviations and generalizations with respect to the very basic case \eqref{sol stazionarie pm} displayed, for the one-dimensional reduction of the problem discussed in Subsection \ref{subsection stationary limit}. Having this in mind, the formal expansion  
\beq
\Phi(\xi)=(\xi  -\xi_0)^{-j}  \sum_{k=k_0}^\infty f_k \,\, (\xi  -\xi_0)^k  \,\,, 
\label{Phi Laurent}
\eeq
being the $f_k$'s constants, can be thence resorted to identify all possible dominant balances, i.e. the singularities whose form behaves like  $\Phi \propto   (\xi - \xi_0)^{-j}$. 
It is readily seen that demanding $j$ to be a positive integer leads to the identification $j=1$, 
so that the singular dominant behavior is associated with a single pole. 
Going further in the analysis, the \emph{resonance} $r=1$ is found (in addition to $r=-1$). So, the two arbitrary constants entering  the Laurent series representation of the solution \eqref{Phi Laurent} with $j=1$   are given by the movable pole $\xi_0$ and the coefficient $f_1$. To our aims, it is significant to focus on the role of dominant term about a singular point $\xi_0$ in the local representations of $\Phi$, 
\beq
\Phi(\xi)\cong \frac{1}{\xi  -\xi_0} \,\,.
\label{Phi dominant xi0}
\eeq
That is, for \eqref{def xi} and assuming $\Delta\neq0$, we can take for Eq. \eqref{veq} the approximate solution 
\beq
v(x,t) \cong  \frac{ \eta \sigma c_1 \, t}{t(c_1 x -c_2 t) -\xi_0^*} = \frac{\eta \sigma}{  x -x_s(t) }
\label{v dominant in}
\eeq
about a singular point (the term $\sigma c_3 /c_1$ can be patently omitted),  being 
$\xi_0^*=\eta^2 c_1^3 \Delta^{-1} \, \xi_0 $ and $x_s(t)$ given by formula \eqref{x(t)} with $\xi/B \to \xi_0^*$, i.e.
\beq
x_s(t)=\frac{1}{c_1} \left(c_2 t +\frac{\xi_0^*}{t}\right) \,\, 
\label{xs(t)}
\eeq
($c_1\neq0$). At any given $t\neq0$, the  singularity shows up when the real variable $x$  attains the finite value $x_s(t)$. When $\xi^*_0=0$, the pole $x_s(t)$  simply depends linearly on $t$ and translates with constant velocity 
$c_2/c_1$. If $\xi_0^*\neq0$  the quantity $x_s$ can be positive or negative  depending on the signs of $\xi_0^*/c_1$ and $c_2/c_1$.  The motion of  poles of the rational solution \eqref{v dominant in} is easily inferred.  We thus see that when $\xi_0^*/c_1$ and $c_2/c_1$ are either both positive or both negative (cases i) and iv) of previous Section) the same value of $x_s$ can be obtained for two distinct values of the local variable $t>0$, meaning that a reversal of the pole motion is exhibited.    When $\xi_0^*/c_1>0$ and $c_2/c_1<0$ or $\xi_0^*/c_1<0$ and $c_2/c_1>0$ -cases ii) and iii)-   there is instead a one-to-one correspondence   based on a monotonic behavior for the function $x_s(t)$. 
Qualitative behavior of $x_s(t)$ when $t>0$ and $\xi^*\neq0$ is as in Fig. \ref{moto poli}.  Also remark that when $\Delta=c_1^2c_4+c_2c_3^2=0$ the coefficients $c_3$ and $c_4$ play no role on the identification and on the motion of the pole $x_s(t)$ as they do not enter in the definition of neither the variable, which now reads  $\xi=t\, (c_1 x-c_2 t)$, nor in defining a normalized value $\xi^*_0$ of the pole $\xi_0$ (since now one  would have \eqref{x(t)} with $\xi\to \xi_0$). 

To gain a picture of how its solutions evolve away from a singularity of the type  \eqref{Phi dominant xi0}, 
a numerical integration of equation (\ref{new EQ riduzione xi}) can be performed. Remark that Equation 
\eqref{new EQ riduzione xi} does not admit invariance neither under reflections $\xi\to-\xi$ nor under translations $\xi\to\xi +\delta \xi$.  

\begin{figure}[h]
\begin{center}
\includegraphics[scale=0.6]{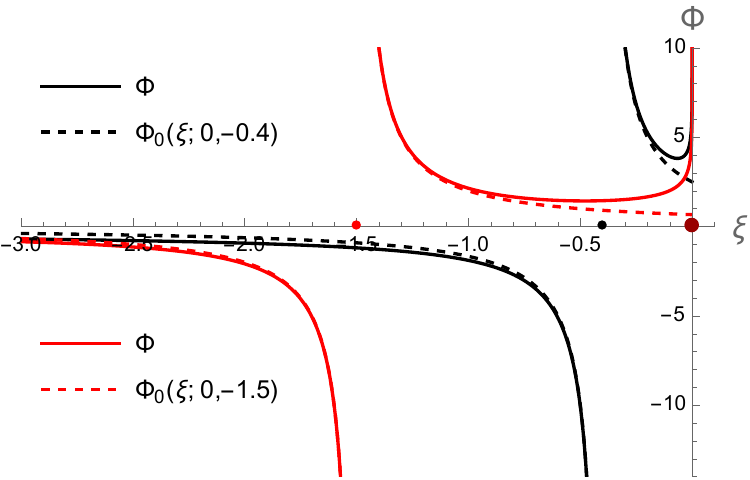}  a) \qquad \qquad
\includegraphics[scale=0.6]{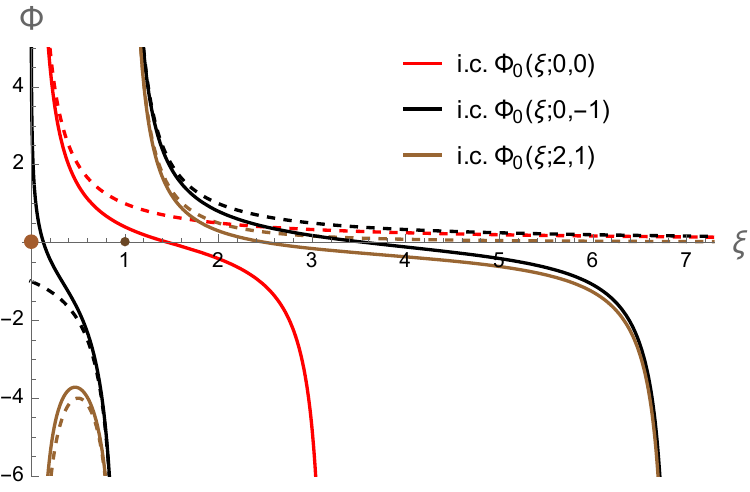}  b)
\caption{Numerical integration of Eq. (\ref{new EQ riduzione xi}) upon superimposition of initial conditions based on 
functions \eqref{sol Phi caso Delta 0} for distinguished choices of parameters. Functions used to set initial conditions are represented by the dashed curves. 
a) Negative simple poles in $\xi_0=-0.4$ (black) and $\xi_0=-1.5$ (red).   
 b) Comparison of curves obtained by superimposing initial conditions either matching simple pole functions at $\xi_0=0,1$ (red and black curves, respectively) or 
\eqref{sol Phi caso Delta 0} with $\alpha_1=2$ and $\alpha_2=1$ (brown) which possess both those singularities.
}
\label{integrazione xs nuove}
\end{center}
\end{figure}

Figure \ref{integrazione xs nuove} pertains solutions $\Phi$ to Equations \eqref{new EQ riduzione xi 2} obtained with initial condition of the simple pole type about $\xi_0<0$. That is, initial conditions have been imposed by assigning that,  at a point very close to a certain $\xi_0>0$, solutions $\Phi$ and $\Phi'$ assume the same values of the function $(\xi-\xi_0)^{-1}$ and of its differential. The initial condition matching takes place very close to singularities and for solutions values rather above the domain considered in plots.  
Integration is performed by increasing the variable $\xi$ from a pole until a successive singularity develops. 
Numerical integration of (\ref{new EQ riduzione xi 2}) in the positive and negative domains of the variable $\xi$ are performed separately. In particular, Figure \ref{integrazione xs nuove}.a) refers to the case $\xi_0<0$. To the left of this singularity, solutions essentially comply with the simple pole function  $(\xi-\xi_0)^{-1}$. The plot also tells how the solution runs to a novel singularity located at the origin. When $\xi$ increases and begins to be distant enough from $\xi_0<0$ the memory of the pole begins to get lost, and a change of convexity  anticipates a growth at a noticeably high rate while approaching the singular point $\xi=0$. Red and black curves in 
Figure \ref{integrazione xs nuove}.b)   report results 
one obtains by integrating Equation (\ref{new EQ riduzione xi 2})  over the domain $\xi>0$ after superimposing  initial conditions of the simple pole type $\Phi\simeq (\xi-\xi_0)^{-1}$, with $\xi_0>0$. After increasing $\xi$ a while, the portion of solutions that develops to the right of the point $\xi_0>0$ are pushed down and directed to a singular behavior, with a  character resembling to $\tan$-type functions. A translation of the singular point $\xi_0$ provokes a dilation of the interval between the initial and the newly formed singularities (see the red and black curves in curves in \ref{integrazione xs nuove}.b)).
Confronted with \eqref{new EQ riduzione xi 1} with the same initial conditions, it can be thus concluded that there is a substantial role played by the last term in Equation \eqref{new EQ riduzione xi 2} that sustains the generation of an additional singularity at the origin $\xi=0$.

\subsection{Sensitivity to initial conditions}

Discussing solutions to Equation (\ref{new EQ riduzione xi 2}) would clearly benefit of taking into account initial conditions other than those used so far. At the beginning of this Section,  we have seen that simple pole behavior can be extracted for solutions to Equation \eqref{new EQ riduzione xi 2}. But we have also learned in Section \ref{section Delta 0} how 
singularities of different types originate for functions $\Phi$ when $\Delta =0$. It makes therefore sense to look at solutions  to Equation \eqref{new EQ riduzione xi 2} based on sampling of initial conditions establishing a more direct connection with general solutions to the problem \eqref{new EQ riduzione xi 1}. 
For instance, we set initial conditions for \eqref{new EQ riduzione xi 2} demanding solution values and their first derivatives as given 
by  the function \eqref{sol Phi caso Delta 0} near a singularity. After doing so, families of numerical solutions to Equation \eqref{new EQ riduzione xi 2} 
follow that can be seen as deformations of the solutions \eqref{sol Phi caso Delta 0},  from which they are expected to inherit some major features.
 Figure \ref{integrazione xs nuove}.b)   considers also what is obtained by this strategy. 
While not a substantial dependence on this new initial condition, compared to the simple pole initial condition,
is displayed to the right of the singularity (see the brown and black $\tan$-type shapes on the right of Figure \ref{integrazione xs nuove}.b)), quite a difference is manifested on its left side. As solid and dotted brown curves in Figure \ref{integrazione xs nuove}.b) show for smaller $\xi$, 
solutions to \eqref{new EQ riduzione xi 2} associated with initial data as given by  $\Phi_0$ tend to stay adherent to $\Phi_0$ itself, thence preserving their same singular behavior approaching the origin. A striking dissimilarity is presented instead in the same domain between the origin and the singularity for the solution.  
Starting from the left of the singularity to the origin, the solution to \eqref{new EQ riduzione xi 2} 
(left solid black curve for small values of $\xi$) leaves the $(\xi-\xi_0)^{-1}$ curve (dotted black) determining its initial simple pole imprinting so to generate again a singularity at the origin, but this time progressing to the positive infinity value through a $\tan$-type profile.

 \begin{figure}[h]
\begin{center}
\includegraphics[scale=0.6]{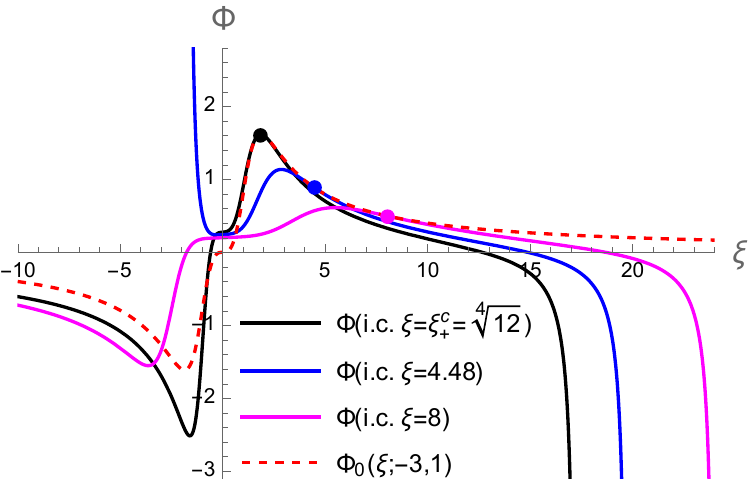}\,\, a)
\includegraphics[scale=0.6]{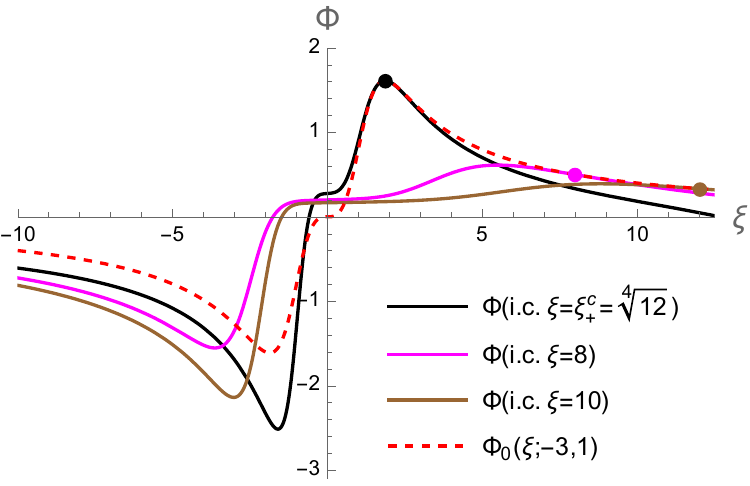}\,\, b)
\caption{Comparison between functions solving (\ref{new EQ riduzione xi 2}) and solution \eqref{sol Phi caso Delta 0} pertinent to  $\Delta=0$.  
 $\Phi(\xi)$ and $\Phi_0(\xi; -3,1)$ are required to pass both through chosen individual points (colored disks) with the same rate.
 The black curve denotes the solution demanded to possess the same maximum  (black disk) of $\Phi_0(\xi; -3,1)$ (red dashed) at 
 $\xi=\xi_{+}^c =  12^{1/4}$.  
a)  Deformation of the reference profile $\Phi_0(\xi; -3,1)$ and sudden apparence of a dissimilar curve (blue). 
b) Dynamics of the minimum by varying initial conditions. Lowering the maximum does not necessarily 
introduce an overall smoothing of $\Phi$ in the $\xi<0$ domain (compare magenta and brown curves).  
\label{integrazione Phi nuove 2}
}
\end{center}
\end{figure}

Having payed attention on consequences of initial conditions near the singularity, devised on the observation of dominant term \eqref{Phi dominant xi0} and solutions to \eqref{new EQ riduzione xi 1}, in order to having a more complete picture on the problem  \eqref{new EQ riduzione xi 2}, it is worth to pursue the same strategy of screening the 
solutions that are isolated  by the enforcement of data similar to that of an assigned reference function \eqref{sol Phi caso Delta 0}.  However, at this time data are  taken at some point far from singularities of $\Phi_0$.  Stationary  points can be chosen  as reference for initial conditions, for instance. 
In Figure \ref{integrazione Phi nuove 2}, in particular, the solution to  \eqref{new EQ riduzione xi 2} is asked to develop the same local maxima  of the function \eqref{sol Phi caso Delta 0} with the off  $\alpha_1=-3$ and $\alpha_2=1$ (red dashed).  A curve emerges that is not anymore anti-symmetric with respect to the vertical axis,  and an asymptote comes after the maximum in a manner analogous to what we have formerly seen to progress from a superimposed vertical asymptote, see Fig. \ref{integrazione xs nuove}.b). To some extend, the $\Phi_0$-pattern is better preserved before the maximum, albeit it is evident that also the anti-symmetry between the maximum and the minimum breaks down. The latter is visibly lowered, like most of the curve in the $\xi<0$ domain. The raising of the function to positive values in the proximity of the origin is also clear, and this contributes to the formation of a greater dip conducting to the minimum.

Graphs of Figure \ref{integrazione Phi nuove 2} also illustrate what can be displayed by still taking $\Phi_0(\xi; -3,1)$ as trial function to settle initial conditions  for the ODE \eqref{new EQ riduzione xi 2}, but choosing  points on it different from the maximum. In particular,  we have considered points to the right of the maximum, where the function smoothly decays monotonically. With the diminishing of $\Phi_0(\xi; -3,1)$ and $\Phi'_0(\xi; -3,1)$ there is a general tendency to push down the maximum for $\Phi$, which moves uniformly to the right, see magenta and brown curves in Figures  \ref{integrazione Phi nuove 2}.a)-b).  A memory of the minimum developing for $\Phi_0(\xi; -3,1)$ is also seen, but in the region 
$\xi<0$ the function $\Phi$ does not experience the same flattening effect taking place for $\xi>0$; see the magenta curve in Figure \ref{integrazione Phi nuove 2}.a). Actually,  Figure  \ref{integrazione Phi nuove 2}.b)  shows that while the right part of the curve $\Phi$ attains lower values for smoother initial conditions (magenta and brown disks), the left part does not: the minimum for the magenta curve  is actually less peaked than that of the brown curve which also tends to get back closer to origin.  Also remark that 
the output profile for $\Phi$ can change abruptly in correspondence of certain initial conditions. This  is shown by means of the blue curve in Fig. \ref{integrazione Phi nuove 2}.a)   
 whose shape evidently reminds the portion of $\Phi_0$ developing on the right of the singularity for even negative $\alpha_1$ and $\alpha_2=1$, but with in addition the newly introduced movable singularity standing out for positive $\xi$.  This can be intuitively understood on the grounds that the choosing of initial conditions for $\Phi$ and $\Phi'$ may in practice act as selecting a $\Phi_0$ with different pairs of $\alpha_1$ and $\alpha_2$.

\subsection{Remarks on solutions  \eqref{def xi} to  \eqref{veq} for $\Delta\neq0$}
 
\begin{figure}[h]
\begin{center}
\includegraphics[scale=0.6]{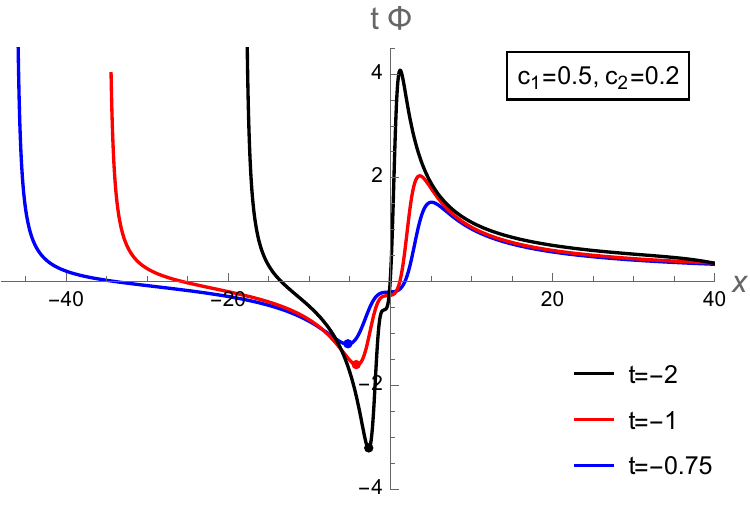} \,\, a) 
\includegraphics[scale=0.6]{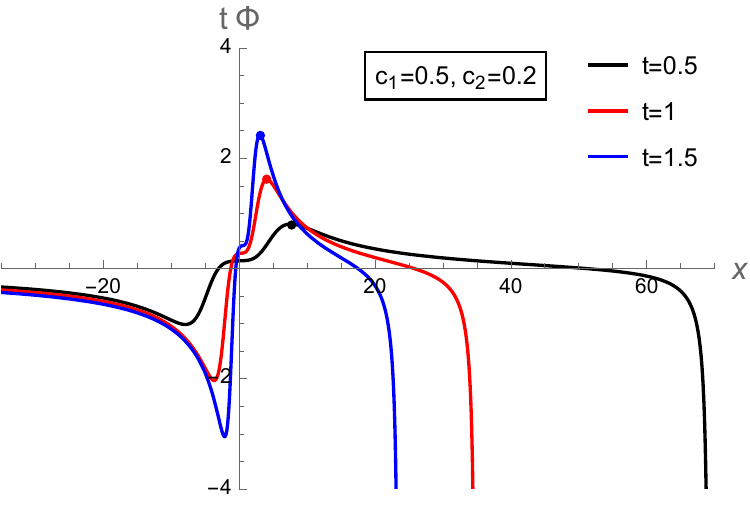}\,\, b)
\caption{   Examples of function $t\Phi(x,t)$ that conseque from \eqref{new EQ riduzione xi 2} at different times and for $c_{1}=0.5$, $c_2=0.2$ and $\Delta=\eta^2 c_1^4$. Colored dots refer to the inital value assignment for the function $\Phi$ coinciding with the maximum of \eqref{sol Phi caso Delta 0} with $\alpha_1=-3$ and $\alpha_2=1$. a) Negative $t$. b) Positive $t$. } 
 \label{plottPhiDeltat}
\end{center}
\end{figure}

 We have previously gained an insight on some peculiarities of a class of solutions to  \eqref{new EQ riduzione xi} and the next step would be 
arguing the pulling back to the original equation \eqref{veq}  acted via  \eqref{def xi}. The analysis of the effects on solutions to \eqref{veq}  can be prospected straight away founded on observations and guiding route put together for the case $\Delta=0$ treated in  Section  \ref{section Delta 0}. We shall not argue therefore the complete casistics for the choices that can be made for the coefficients $c_j$ for it is very intelligible from \eqref{def xi}. In Figure \ref{plottPhiDeltat}, we therefore produce only the function $t\Phi$ determining solutions to \eqref{veq} from the solution to  (\ref{new EQ riduzione xi 2}) drawn in Figure \ref{integrazione Phi nuove 2}. The predicted flattening out during the evolution near $t=0$, the exchange with reflections between the quadrants of the curve sections after the  annulment at $t=0$ and the amplification of the peaks in the later evolution are manifest.  
We conclude the Section by emphasizing that in potential applicative problems relying on equation \eqref{veq} it may be also necessary to 
 ascertain the consistency with possible constraint on domains of variables for the specific subject matter under study. For instance, if $v$ is definite positive, those portions of the $\Phi$ curves have to be selected such that also the condition $v\ge 0$ comes true.  Depending on the signs of structural coefficients $c_j$'s, in turn this demands  taking into account either the condition $B \eta \frac{c_1^2 }{c_3} t \Phi<1$  or the condition $B \eta \frac{c_1^2 }{c_3} t \Phi>1$, where $B=1$ if $\Delta=0$ or $B= \Delta \eta^{-2} c_1^{-4}$ otherwise (Eq. \eqref{B caso 2}); i.e., either $ \eta \frac{c_1^2 }{c_3} t \Phi>1$ or $ \eta \frac{c_1^2 }{c_3} t \Phi<1$ for $\Delta=0$, and either
$  \frac{  \Delta }{\eta c_1^2 c_3}  t \Phi>1$  or $  \frac{  \Delta }{\eta c_1^2 c_3} t \Phi>1$ if $\Delta\neq0$.

 \section{N-pole dynamics of rational solutions}
 \label{section poles}
 
 By looking at the Equation \eqref{veq} from the angle of local symmetry properties and similarity solutions,   we have   gained through the reduction \eqref{def xi} an interesting perspective on some pivotal characteristics owned by its solutions. The study we presented is clearly of limited extend, and the investigation of other fundamental features is in order. Among the open problems there is, for instance, the  comprehension of aspects such as the existence of other distinctive classes of solutions and 
 the possibility to proceed with their classification. In particular, shedding light on the 
 existence of rational solutions comprising multiple simple poles seems to be a naturally due development. 
Indeed, singular solutions to remarkable integrable equations and their pole dynamics have proved to be intriguingly linked to the dynamics of particles in many-body systems   (see, for instance, the seminal papers \cite{airault,Choodnovsky,calogero 2}) as well as to rogue waves (see \cite{dubard2, gaillard,clarkson2}), and are still subject of active investigations about the rational solutions in the KP hierarchy  and Painlevé equations \cite{zabrodin2019,zabrodin2020,zabrodin2021, clarkson3}. 
  In addressing this issue, guidelines can be taken from discussions concerning other integrable equations, as put forward for instance in \cite{Choodnovsky}.
 The most natural connection in this respect clearly is seized with the standard Bateman-Burgers equation. In such a case, the existence of rational solutions and the analysis of their poles properties has been nicely 
 investigated in \cite{deconinck}, for instance.  By proceeding similarly to  \cite{Choodnovsky,deconinck},  the $N$-pole ansatz 
 \beq
 v(x,t)= \eta \sigma \sum_{k=1}^N \frac{R_k(t)}{x-x_k(t)}
 \label{v N-poles}
 \eeq 
 can be substitued in \eqref{veq}, and by later setting $x=x_j+\epsilon$, the resulting equation can be expanded in powers of $\epsilon$. As in the Bateman-Burgers case, equating the most singular terms prompts to $R_k(t)=1$. We observe here, however, that quite a more complex  pole dynamics is suggested for rational solution of Equation  \eqref{veq}, at least when $\Delta\neq0$. Indeed, by collecting terms at the successive order, one finds 
 that the motion of poles $x_k(t)$ superimposed via \eqref{v N-poles} can be thus ascribed, at this level of approximation, to the concurrence of two  mechanisms: a translation at fixed constant speed unavoidably entering in the matter once $c_1c_4\neq0$, and an additional motion obeying a nontrivial differential system. That is:
 \beq
 x_k(t)=-\frac{c_1 c_4}{c_3^3} \, t + y_k(t)= \left( \frac{c_1}{c_2}-\frac{\Delta }{c_1c_3^2} \right) t + y_k(t)\,\, , 
 \label{x_k y_k}
 \eeq
 where the functions $y_j(t)$ are determined by solutions of the following dynamical system:
 \begin{align}
 	\dot{y}_k(t)=&
 	\frac{\Delta}{c_3^3} \sum_{m=1}^{N-1}   (m+1)! \left(-\frac{c_1}{c_3}\right)^{m-1}\,  \eta^m \, 
 	{\cal P} ^{(m)}_k\left[\{y_m\}_1^N\right] \,\, , 	 
 	\qquad \qquad \left( \dot{ }  =\frac{d}{dt} \right) \,\, , 
 	\label{sistema poli IEW}
 \end{align}
 the $ {\cal P} ^{(m)}_k$'s denoting  the sum of products of distinct terms $(y_k-y_j)^{-1}$, i.e. 
 \beq
 {\cal P} ^{(1)}_k \left[ \{y_m\}_1^N \right]=
 \sum\limits_{\substack{j=1 \\ i\neq k}}^N \frac{1}{y_k-y_j} \,\, ,
 \qquad {\cal P} ^{(2)}_k\left[\{y_m\}_1^N\right]=  \sum\limits_{\substack{j,m=1 \\ j,m\neq k \\ j \neq m}} ^N  \frac{1}{(y_k-y_j) (y_k-y_m)} \,\, , 
 \eeq 
 \beq
 {\cal P} ^{(3)}_k\left[\{y_m\}_1^N\right]=  \sum\limits_{\substack{j,m,r=1 \\  j,m,r\neq k \\  j\neq m, j\neq r, m\neq  r}}^N
 \frac{1}{(y_k-y_j)(y_k-y_m)(y_k-y_r)}
 \label{sistema poli IEW 2}
 \eeq 
 and so forth. The situation is therefore much different from the standard Bateman-Burgers equation with diffusivity $\eta$ 
 issuing for $c_1\to 0$ (along with the normalization conditions $c_3=-c_2$ and $\sigma=-2$), for which 
 the motion of poles entering a N-pole ansatz of the type \eqref{v N-poles} would be determined by the hamiltonian differential system 
 $ \dot{x}_k(t)=-2\eta {\cal P} ^{(1)}_k \left[ \{x_m\}_1^N \right] $. 
 Then, there is an interaction among poles that is not limited to just pairwise couplings as it happens in the standard Bateman-Burgers equation  \cite{deconinck}. 
 The manner in which a pole $x_k(t)$ in \eqref{v N-poles} varies  is ruled by its correlations with an increasing number of the other distinct poles,  
 until all the poles appear. Two remarks are clearly in order on this statement. First, it has been assumed that
 $\Delta=c_1^2c_4+c_2c_3^2\neq0$. When $\Delta=0$ one is merely lead instead to $t$-linear translations of poles with rate  $\dot{x}_k(t)=\frac{c_2}{c_1} t$. 
 Secondly, under the circumstance that $\eta$ is a small perturbation parameter, one may reason about disregarding higher-order terms in powers of $\eta$. By maintaining only the lowest order term in the right hand side of \eqref{sistema poli IEW},  the dynamical system governing the contribution $y_k(t)$ to the motion of poles $x_k(t)$ 
 shares essentially the same form found in the Bateman-Burgers case:
 \beq
 \dot{y}_k(t)=2\frac{\Delta}{c_3^3} \eta \, {\cal P} ^{(1)}_k \left[ \{y_m\}_1^N \right]\,\, . \eeq
 The investigation of the differential system \eqref{sistema poli IEW}  goes beyond our current scopes. We limit ourselves to remark that complete integrability is expected to be evincible using the Cole-Hopf transformation and the relationship between equations \eqref{veq} and \eqref{phieq}.

\section{Discussion and conclusions}
\label{section conclusions}

In this communication, we have addressed the study of the nonlinear 1+1 dimensional partial differential equation \eqref{veq}, i.e. 
  \begin{equation}
\label{veq2}
\partial_t v + \partial_{x} \left\{ \frac{1}{c_{1} v + c_{3} \sigma} \left[c_{2} v^{2} + c_{4} \sigma^{2} + \sigma \eta \left(c_{1} \partial_{t}v + c_{2} \partial_{x} v \right) \right] \right\} = 0 \,\,, 
\nonumber
\end{equation}
representing a non-evolutive generalisation of known diffusive/dissipative equations. 
The equation has been introduced recently in \cite{giglio} and ensues from rather general grounds. Indeed, it  can be obtained from a 1+1 differential conservation law where: {\it i)}
 the flux density depends both on the density and (linearly) from the first derivatives of density with respect to the local variables; 
 {\it ii)}  the linearisability via a Cole-Hopf transform is demanded in addition. 
To date, very little is known on Equation \eqref{veq}  and its solutions. When one or more of the coefficients $c_j$ do vanish, some crucial simplifications may occur that enable to determine solutions almost effortlessy, as we commented in Subsection \ref{subsection cj 0}. In particular, the simple $\tan$-type \eqref{sol tan x} or rational solutions
\eqref{sol stazionarie pm} are admitted. The circumstance definitively motivates the paying attention on the occurrence of singularities for solutions to \eqref{veq}.

In the present communication, by performing a similarity reduction dictated by one of the local symmetry generators underlying the equation \eqref{veq}, 
 we have shown that a nonlinear ordinary differential equation arises which is connected to the  Painlev\'e III equation. While standard arguments concerning the existence of solutions with simple poles prove their usefulness, the discussion we presented here in this regard took quite a benefit from the circumstance 
that the model is exactly solvable   in the special case  where the constraint $\Delta=c_1^2c_4+c_2c_3^2=0$ do hold. The general real solution thence originated takes a simple mathematical structure with a still rich dynamics and interesting key-elements, e.g. in respect to the presence of singularities how we have detailed.  Regardless the potential marginal relevance of such a strict binding $\Delta =0$ in concrete applications to specific problems, the circumstance of exact solvability provides indeed first remarkable hints concerning what to expect for solutions to the general differential problem when $\Delta\neq0$, at least within fairly identified assumptions and regimes. This allows for gaining a clue also on the unavoidable qualitative differences,  such as the generation of additional divergences, as we argued in Section \ref{section Delta gen} where the strategy of selecting solutions by tailoring them to an assigned solution pertaining $\Delta=0$    has been pursued.  We have been also attentive to the implications in respect to the $t$-dependent dynamics of poles and asymptotes for the solution, showing an inversion of their motions. These results put a basis for the understanding of the general case $\Delta\neq0$, for which a more singular  behavior has been shown to emerge.

Future investigations concerned with \eqref{veq} are  definitively in order and  they may naturally include both other mathematical  issues and more 
applicative problems. We already understood in Section \ref{section poles} that a deeper comprehension of the integrability properties underlying the dynamics of multiple poles would be desirable, in a manner parallel to what has been done for the Bateman-Burgers equation for instance \cite{deconinck}.
Another  elucidation looks to be  in respect to the identification of non-local symmetries. Besides providing insights on the possibility to have receipts for transferring possible analytical results,  hints on the settling of diverse coordinate transformations may play a role as regards aspects like the reduction to normal form, classification of solutions and characterization of possible correlated hierarchies  in a way similar to the case of the non-evolutionary viscous scalar reduction of the two-component Camassa-Holm equation considered in \cite{arsie}. Generalised symmetries determining Miura-type actions play indeed a decisive role in that framework.  It would be also interesting to understand if they may enable to connect \eqref{veq} to a Painlev\'e equation  in the general case when $\Delta\neq0$, as we have seen that for $\Delta=0$ the equation is related to the Painlev\'e III  through a Cole-Hopf type transformation. 
 Other developments may be more focused on the application of Equation \eqref{veq}, including those for purposes potentially different from the fluid systems considered so far, such as in the context of complex systems based on  mean-field spin models. The idea of formally describing the governing behavior of relevant statistical quantities through the solutions of PDEs is starting to be appraised even for problems in  reaction kinetics \cite{agliari2018} and artificial intelligence \cite{agliari2022}, for instance. This may raise the question about the possible need to  implement the model \eqref{internal energy expansion} while dealing with specific problems, such as by taking into account other non-local terms in the expansion that may cure or smoothen appearing criticalities and allow for the identification of multiple scales of both mathematical and applicative relevance. We finally mention that natural advances  of the investigation performed in this work entail   multi-component integrable conservation laws, for instance those arising in the treatment of nematic liquid crystal systems \cite{dematteis,dematteis2}. Extension of our approach to   integrable hydrodynamic chains  may also prove to be insightful, e.g. in the context of random matrix models \cite{benassi}.

\appendix

\section{Proofs of Proposition \ref{stationary points}, Proposition \ref{poles rational} and Proposition \ref{stationary points rational}}
\label{appendice 2}

\setcounter{theorem}{1}
\begin{proposition}[Stationary points  of solution~\eqref{sol Phi caso Delta 0} for integer values of $\alpha_1$]
	Let $\alpha_1 \in \mathbb{Z}$, $\alpha_2>0$, $\xi^{(0)} =0$ and  $\xi_{\pm}^c =\pm |  \alpha_1 (1-\alpha_1)\alpha_2|^{\frac{1}{1-\alpha_1 }}$.  The stationary points of solution \eqref{sol Phi caso Delta 0}   are listed below.
	
	\begin{enumerate}[i)]
		\item  Three stationary points located at $\xi=\xi^{(0)}$ (inflection point) and  $\xi=\xi_{\pm}^c$ for $\alpha_1$ odd  negative integer with $\alpha_1<-1$.
				\item  Two stationary points when $\alpha_1=-1$, located at $\xi=\xi_{\pm}^c$ .
		\item  Two stationary points for $\alpha_1$ even negative integer,  located at $\xi=\xi^{(0)}$ and $\xi=\xi_{+}^c$.  
		\item One single stationary point located at $\xi=\xi_+^c$  for $\alpha_1$ even positive integer.
		\item Two stationary points  for odd positive integers $\alpha_1$ located  at $\xi=\xi_{\pm}^c$.
	\end{enumerate}
\indent
	\begin{proof}
		The proof is based on similar arguments used in Proposition~\ref{poles}. The stationary points of \eqref{sol Phi caso Delta 0} are given by  the zeros of $\Phi_0 ' (\xi; \alpha_1,\alpha_2)=(\alpha_1-1) \frac{\left[ \alpha_1(\alpha_1-1)\alpha_2 \xi^{\alpha_1-1} -1\right]}{\left[( \alpha_1-1)\alpha_2\xi^{\alpha_1} - \xi \right]^2} $. \\
		Let us first consider the case $\alpha_1\in \mathbb{Z}_{<0}$. The stationary points are given by the real roots of the polynomial $ \mathcal{\tilde{P}}^{-} (\xi):= \left[ \alpha_1(\alpha_1-1)\alpha_2 - \xi^{1-\alpha_1} \right] \xi^{\alpha_1-1}$. One root  $\xi^{(0)}=0$ arises for $\alpha_1 <-1$. One can verify that $\Phi_0''(\xi^{(0)})=0$, hence giving an inflection point. The other roots for negative integers are given by $\xi_l=|  \alpha_1 (1-\alpha_1)\alpha_2 |^{\frac{1}{1-\alpha_1}}  e^{i\frac{ 2 l \pi}{1- \alpha_1}}$, $l=0, 1, \cdots, -\alpha_1$.  The real solution $\xi_+^c=|  \alpha_1 (1-\alpha_1) \alpha_2 | ^{\frac{1}{1-\alpha _1}}$ is obtained when $l=0$, while  $\xi_-^c=-|  \alpha_1 (1-\alpha_1) \alpha_2 | ^{\frac{1}{1-\alpha _1}}$ is obtained for $\alpha_1$ odd  when $l=\frac{1-\alpha_1}{2}$. These prove i), ii) and iii). \\
		Let us now consider the case $\alpha_1 \in \mathbb{Z}_{>0}$. The case $\alpha_1=1$ identifies the null solution as discussed earlier. When $\alpha_1>1$ we have that stationary points are real roots of the polynomial $ \mathcal{\tilde{P}}^{+} (\xi):=  \alpha_1(\alpha_1-1)\alpha_2 \xi^{\alpha_1-1}- 1 $,  $\xi_l=|  \alpha_1 (1-\alpha_1)\alpha_2 |^{\frac{1}{1-\alpha_1}}  e^{i\frac{ 2 l \pi}{1-\alpha_1}}$, $l=0, 1, \cdots, \alpha_1-2$. We promptly identify the real root $\xi_+^c=|  \alpha_1 (1-\alpha_1)\alpha_2 |^{\frac{1}{1-\alpha_1}} $. A second real solution occurs for odd integers only, requiring  $2 l \pi=(1-\alpha_1)\pi$, that is $l=\frac{1-\alpha_1}{2}$, giving $\xi_-^c=-|  \alpha_1 (1-\alpha_1)\alpha_2 |^{\frac{1}{1-\alpha_1}}$. These complete the proof of iv) and v).
	
	\end{proof}
\end{proposition}

\begin{proposition}[Singularities of \eqref{sol Phi caso Delta 0} for rational values of $\alpha_1$] Let $\alpha_1=\frac{q}{p}$ with $q, p \in \mathbb{Z}$ and $\gcd(q,p)=1$, and consider  $\xi^{(0)}$, $\xi_{\pm}$ defined as in Proposition~\ref{poles}. The real  poles of  \eqref{sol Phi caso Delta 0} are listed below.
	\begin{enumerate}[i)]
				\item Case $\alpha_1<0$.
		\begin{enumerate}[a)]
			\item One singularity located  at $\xi=\xi_-$  (branch point) if $p$ is odd and $q$ is even.
			\item No singularities if $p$ and  $q$ are odd.
			\item One singularity at $\xi=\xi_+$ (branch point) if $p$ is even.  
		\end{enumerate}
		\item Case 
		$0<\alpha_1<1$.  
		\begin{enumerate}[a)]
			\item Two singularities if $p$ is odd and $q$ is even located  at $\xi=\xi^{(0)}$ (branch point) and  $\xi=\xi_{-}$ (branch point).
			\item One singularity   located  at $\xi=\xi^{(0)}$ (branch point) if $p$ and  $q$ are odd.
			\item One singularity located at $\xi=\xi^{(0)}$ (branch point) if $p$ is even.
		\end{enumerate}
	\item Case $\alpha_1>1$.
\begin{enumerate}[a)]
	\item Two singularities if $p$ is odd and $q$ is even located  at $\xi=\xi^{(0)}$ (pole) and  $\xi=\xi_{+}$ (branch point).
	\item Three singularities if $p$ and  $q$ are odd  located  at $\xi=\xi^{(0)}$ (pole) and $\xi=\xi_{\pm}$ (branch points).
	\item Two singularities  at $\xi=\xi^{(0)}$ (pole) and $\xi=\xi_+$ if $p$ (branch point) is even.
\end{enumerate}
		\end{enumerate}
		\begin{proof}
			Wlog, let  $p \in \mathbb{Z}_>0$ and consider the above different cases as $q$ varies in $\mathbb{Z}$. Note that for $p$ even, the solution \eqref{sol Phi caso Delta 0} is defined for $\xi \geq 0$. Also notice that the additional condition $\gcd(q,p)=1$ ensures that to each $\alpha_1 \in \mathbb{Q}$ we associate unambiguously a pair of integers $(q,p)$. \\
			  We will prove  case iii) as an example, $\alpha_1>1$. In this case, singularities are  given by zeros of the polynomial  $\mathcal{P}^{+} (\xi):=  \xi \left[ \left(  (\alpha_1 -1) \alpha_2 \right) \xi^{\frac{q-p}{p}}-1 \right]$. Clearly a singularity occurs at $\xi^{(0)}=0$. Other real solutions are among the following $\xi_l=|  \alpha_1 (1-\alpha_1)\alpha_2 |^{\frac{1}{1-\alpha_1}}  e^{i p\frac{ 2 l \pi}{q-p}}$, $l=0, 1, \cdots,q-p -1$. Real roots are obtained in two cases, either looking for $l$ fulfilling $2 p l = 2k (q-p)$ or  $2 p l = (2k+1) (q-p)$ for some integer $k$. The former gives $\xi=\xi_+$, while the latter gives  $\xi=\xi_-$.  The parity of $q$ and $p$ implies the qualitative scenario. Indeed, $\xi=\xi_+$ arises for all parities of $q$ and $p$, but the situation is different for $\xi=\xi_-$. Indeed, only when $q$ and $p$ are both odd (hence $q-p$ even), an additional solution is found at $\xi=\xi_-$. This proves i). \\
			  Cases i) and ii) are derived following similar procedure and their proof is here omitted. In general, one has to identify the polynomial whose zero correspond to singularities of $\phi_0$, find the complex roots and select the real ones by looking at the parity of $q$ and $p$.
			  
		\end{proof}	
	\end{proposition}

\begin{proposition}[Stationary points of \eqref{sol Phi caso Delta 0} for rational values of $\alpha_1$] Let $\alpha_1=\frac{q}{p}$ with $q, p \in \mathbb{Z}$ and $\gcd(q,p)=1$, and consider  $\xi^{(0)}$, $\xi_{\pm}^c$ defined as in Proposition~\ref{stationary points}. The stationary points of  \eqref{sol Phi caso Delta 0} are listed below.
	\begin{enumerate}[i)]
		\item Case $\alpha_1<0$.
		\begin{enumerate}[a)]
			\item Two stationary points located  at $\xi=\xi^{(0)}$ and  $\xi=\xi_+^{c}$  if $p$ is odd and $q$ is even.
			\item Three stationary points located at $\xi=\xi^{(0)}$   (inflection point) and  $\xi=\xi_{\pm}^{c}$  if $p$ and  $q$ are odd.
			\item Two stationary points located  at $\xi=\xi^{(0)}$ and  $\xi=\xi_+^{c}$  if $p$ is even.  
		\end{enumerate}
		\item Case
		$0<\alpha_1<1$.  
		\begin{enumerate}[a)]
			\item One stationary point at $\xi=\xi_-^{c}$ if $p$ is odd and $q$ is even.
			\item No stationary points  if $p$ and  $q$ are odd.
			\item No stationary points  if $p$ is even\footnote{If one considers $\xi^p=-\sqrt[p]{\xi}$ a stationary point is located at $\xi=\xi_+^{c}$.}.
		\end{enumerate}
		\item Case $\alpha_1>1$.
		\begin{enumerate}[a)]
			\item One stationary point located at  $\xi=\xi_+^{c}$  if $p$ is odd and $q$ is even.
			\item Two stationary points   located  at $\xi=\xi_{\pm}^c$  if $p$ and  $q$ are odd.
			\item One stationary point at $\xi=\xi_+^c$  if $p$  is even.
		\end{enumerate}
	\end{enumerate}
	\begin{proof}
Let  $p \in \mathbb{Z}_>0$ and consider the above different cases as $q$ varies in $\mathbb{Z}$. We will prove case i) as an example, $\alpha_1<0$.	 
In this case, stationary points are  real zeros of the polynomial 
$P^{-}({\xi})=\xi^{-2 \alpha_1} \left[ \alpha_1\left(\alpha_1-1)\alpha_2 -\xi^{1-\alpha_1}\right) \right]$. 
Clearly, $\xi=0$ is always a stationary point. The other stationary points are the real roots among $\xi_l=|\alpha_1(\alpha_1-1)\alpha_2|^{\frac{1}{1-\alpha_1}} e^{i\frac{ 2 \pi l p}{p-q} }$, $l=0,1, \cdots, p-q-1$. 
Real roots are obtained in two cases, either looking for $l$ fulfilling $2 p l = 2k (q-p)$ or  $2 p l = (2k+1) (q-p)$ for some integer $k$. The former would imply 
$\xi=\xi_+^{c}$, while the latter would lead to  $\xi=\xi_-^{c}$. A second stationary point located at $\xi=\xi_+^c$ arises for all parities of $q$ and $p$. 
A third stationary point located at $\xi=\xi_-^c$ arises when $p$ and $q$ are odd only, hence proving i). Cases ii) and iii) are derived following similar procedure and their proof is here omitted. 
	\end{proof}	
\end{proposition}

    \acknowledgments
F.G. and G.L. would like to thank the {\em Isaac Newton Institute for Mathematical Sciences}, Cambridge, for support and hospitality during the programme \emph{Dispersive hydrodynamics: mathematics, simulation and experiments, with applications in nonlinear waves} where work on this paper was undertaken. This work was supported by EPSRC grant no EP/R014604/1. F.G. also acknowledges the hospitality of the  Lecce's division of I.N.F.N. and of the {\em Department of Mathematics and Physics ``Ennio De Giorgi"} of the University of Salento.
 G.L. also acknowledges the hospitality of the {\em School of Mathematics and Statistics} of the University of Glasgow. 
 G.L. and L.M. are partially supported by INFN IS-MMNLP.  Authors are indebted with A. Moro for useful discussions.
 G.L. thanks F. Curto for suggestions leading to the improvement of the figures. 
 
 For the purpose of open access, the authors have applied a Creative Commons Attribution (CC BY) licence to any Author 
 Accepted Manuscript version arising from this submission.

\end{document}